\documentclass[10pt,final,3p, nonatbib]{elsarticle}

\newcommand{\textsave}[1]{}

\usepackage{physics}

\usepackage[dvipsnames]{xcolor}
\definecolor{darkgreen}{RGB}{0,100,0}

\usepackage[normalem]{ulem}

\usepackage[utf8]{inputenc} 
\usepackage[english]{babel}
\usepackage{csquotes}

\usepackage[
    backend=biber,
    style=numeric,
    sorting=none,
    maxbibnames=999,
    natbib=true
]{biblatex}
\newcommand{\commentMV}[1]{\textcolor{blue}{#1}}

\usepackage{lmodern}
\usepackage{soul}

\usepackage{graphics}
\usepackage{graphicx}

\usepackage{booktabs}
\usepackage{multirow}
\usepackage{array}
\usepackage{amsmath}

\usepackage{amsthm}
\usepackage{verbatim}
\usepackage{relsize}
\usepackage{array}

\usepackage{caption}

\usepackage{algorithm}
\usepackage{algpseudocode}

\usepackage{hyperref}
\newcommand{\doi}[1]{\textsc{doi}: \href{http://dx.doi.org/#1}{\nolinkurl{#1}}}

\definecolor{comment}{rgb}{0.0, 0.5, 0.0}

\hypersetup{
    colorlinks=true,
    linkcolor=blue,
    filecolor=magenta,
    urlcolor=cyan,
}
\usepackage{booktabs}
\usepackage{color}

\definecolor{mygreen}{rgb}{0,0.6,0}
\definecolor{mygray}{rgb}{0.5,0.5,0.5}
\definecolor{mymauve}{rgb}{0.58,0,0.82}

\usepackage{array}
\usepackage{mathtools}
\usepackage{verbatim}
\usepackage{relsize}
\usepackage{multirow}
\usepackage{multicol}

\newcommand{\lstin}[1]{{\ttfamily #1}}
\usepackage{listings}
\usepackage[bitstream-charter]{mathdesign}
\usepackage[T1]{fontenc}
\usepackage{bm}

\DeclareMathAlphabet{\pazocal}{OMS}{zplm}{m}{n}

\usepackage{lscape}		
\usepackage{fancyvrb}

\usepackage[mathscr]{euscript}

\usepackage{textcomp}

\def \bbeta  {{\bm{\beta}}}

\def \x  {\bm{x}} 

\usepackage{bigints}

\usepackage[switch]{lineno}

\newcounter{MVremark}

  \usepackage{lineno}

\journal{Computer Methods in Applied Mechanics and Engineering}

\begin{document}

\begin{frontmatter}



\title{Correlation-aware sequential sampling for multi-output polynomial chaos expansions}

\author{Qitian Lu}  \ead{Qitian.Lu@vutbr.cz}

\author{Miroslav Vo{\v{r}}echovsk{\'y}}\ead{miroslav.vorechovsky@vut.cz}

\author{Luk{\'a}{\v s} Nov{\'a}k\texorpdfstring{\corref{cor1}}{}}  \ead{Lukas.Novak4@vut.cz}  \cortext[cor1]{Corresponding author}
\address{Brno University of Technology, Brno, Czech Republic}

\begin{abstract}
The interaction of multiple physical processes in complex engineering systems is often characterized by several quantities of interest (QoIs) with non-negligible cross-correlations. Accurate uncertainty quantification (UQ) of these responses requires repeated model evaluations, which can be computationally expensive when high-fidelity simulations are involved. As an alternative, polynomial chaos expansion (PCE) provides an efficient surrogate modeling framework for reducing this computational cost owing to its natural structure for statistical analysis, while adaptive sequential sampling can further improve sampling efficiency under a limited model evaluation budget. However, adaptive sampling strategies specifically designed for multi-output PCE remain limited, and existing approaches commonly aggregate information from individual outputs without explicitly accounting for their statistical dependence. When strong correlations exist among QoIs, output-wise aggregation can give greater combined influence to response patterns shared by several outputs. To reduce the influence of such repeated representation on sample allocation, this study develops a principal component analysis-based extension of the multi-output $\Theta$ criterion for adaptive construction of PCE. At each enrichment stage, principal component analysis (PCA) is applied to the standardized model responses evaluated at the current experimental design. The resulting components are uncorrelated with respect to the empirical distribution of that design. The sampling criterion combines component-wise normalization with the retention of selected principal directions, thereby adjusting the relative influence of shared response patterns while preserving distance-based exploration. An adaptive weighting based on the concentration of variance contributions further adjusts this balance to promote broader coverage of the input domain in problems with localized response features. The method is verified through numerical examples with different levels of output coupling and is compared with the conventional multi-output $\Theta$ criterion and reference sampling strategies. The results demonstrate that accounting for output dependence improves the allocation of new samples among multiple response features and enhances the efficiency of multi-output PCE surrogate construction, particularly for strongly correlated QoIs.
\end{abstract}


\begin{highlights}

\item A correlation-aware PCA extension of the multi-output $\Theta$ criterion is proposed.

\item PCA reduces the repeated influence of response features shared by several outputs.

\item Retained components guide sampling, and the final PCEs predict the original outputs.

\item Variance concentration determines the weight of exploitation relative to exploration.

\item Tests show accuracy gains and lower sampling costs in the examined examples.

\end{highlights}


\end{frontmatter}


\section{Introduction}
\label{Introduction}

Uncertainty quantification (UQ) plays an essential role in the analysis and design of engineering systems whose responses are affected by uncertain input parameters \cite{Ghanem2017Handbook}. In many practical applications, system behavior cannot be adequately described by a single scalar response. Instead, multiple quantities of interest (QoIs) are required to characterize different aspects of the same system, such as responses at different spatial locations or time instants, or different physical quantities from coupled multiphysics systems \cite{conti2010bayesian, chen2013flexible}. These QoIs are usually evaluated under the same uncertain inputs and are often governed by related physical mechanisms. Consequently, these outputs are not independent but exhibit different levels of statistical dependence. This dependence is especially relevant for multi-output modeling, where the response set may contain both shared information and specific features to each output.

For high-fidelity numerical simulations, direct uncertainty propagation generally requires a large number of repeated model evaluations, which can become computationally prohibitive when each simulation is expensive (e.g., large-scale models and nonlinear finite element simulations). Surrogate modeling is therefore widely used to reduce the cost of UQ. One of the widely adopted techniques, polynomial chaos expansion (PCE) \cite{Wiener_PCE}, represents the model response using an orthogonal polynomial basis associated with the input probability distributions \cite{xiu2002wiener}. Once a standard PCE is constructed, the mean and variance of the response can be calculated directly from its expansion coefficients, while response probability distributions and sensitivity indices can also be efficiently estimated, making PCE particularly convenient for UQ \cite{sudret2008global, NOVAK2022106808}. However, the performance of PCE also strongly depends on the quality of the training samples used to determine the expansion coefficients \cite{fajraoui2017sequential, HADIGOL2018382}.

Generally, experimental design (ED) strategies can be classified into two categories: non-adaptive (static) designs and adaptive designs \cite{garud2017design}. Static approaches determine the entire ED before any model responses are observed. Among these, Monte Carlo (MC) sampling \cite{metropolis1949monte}, Latin hypercube sampling (LHS) \cite{mckay2000comparison}, and many other space-filling strategies \cite{morris1995exploratory} are commonly employed because they are straightforward and can provide broad coverage of the input domain. When only limited prior knowledge of the underlying response is available, these methods can be a suitable choice. However, their main limitation is that the ED remains unchanged after information on the response becomes available. Consequently, although they may provide uniform coverage of the input space, such coverage does not necessarily ensure that important response features are adequately captured. Under a restricted evaluation budget, samples may be assigned uniformly to smooth or already well-resolved regions, while localized nonlinearities, rapid variations, or poorly approximated regions remain insufficiently represented \cite{fajraoui2017sequential, novak2021variance}. To address this limitation, sequential adaptive design (also referenced as active learning) is proposed, which alternates between surrogate construction and ED enrichment \cite{crombecq2011novel,fuhg2021state}. Specifically, a sampling criterion is evaluated over a set of candidate points using information extracted from the current surrogate. The selected most informative candidates are then evaluated using the original computational model, and the resulting samples are added to the current ED before the surrogate is updated. Active learning criteria are commonly interpreted in terms of exploration and exploitation, where exploration seeks insufficiently sampled regions by various geometric criteria, and exploitation often uses surrogate-informed indicators, such as cross-validation errors or local response variations, to refine regions that are expected to be difficult to approximate. Hybrid strategies combine these two objectives to improve global surrogate accuracy while limiting excessive sample clustering \cite{crombecq2011novel,fuhg2021state}. For regression-based PCE, adaptive ED methods have been developed along several directions. One class of methods focused on sequential optimal designs based on information-matrix criteria~\cite{fajraoui2017sequential}, whereas another combines polynomial basis adaptation with sequential ED enrichment. For example, coherence-optimal sampling supports the sequential enrichment of the ED as the anisotropic polynomial basis evolves \cite{hampton2018basis}, while a weighted least-squares framework coordinates adaptive basis growth and refinement with sequential optimal sampling \cite{thapa2020adaptive}. For reliability analysis, active learning methods of PCE employ bootstrap-based local error estimates \cite{marelli2018active} or composite learning functions \cite{zhou2020active} to direct new samples toward predicted limit state surfaces. Another perspective is represented by the recently proposed $\Theta$ criterion for PCE combining variance-based exploitation with distance-based exploration \cite{novak2021variance}.

For problems involving multiple QoIs, the adaptive sampling criterion must be extended to a vector-valued outputs. When each model evaluation returns all QoIs simultaneously, a straightforward strategy is to construct a sampling measure specific to each output and aggregate the resulting measures into a unified criterion. For example, the multi-output CV-Voronoi method combines weighted-sum and extreme terms to aggregate error estimates across responses \cite{liu2018multi}, while the Jackknife-maximin approach combines Jackknife cross-validation variance with a minimum-distance measure and uses adaptive weights to aggregate the criteria  \cite{maxim2019efficient}. A potential limitation of such aggregation is that it does not explicitly account for output dependence and may therefore assign disproportionate importance to information shared by strongly correlated responses. This issue is particularly important for spatial or temporal fields, whose numerous output components may be governed by a much smaller number of underlying modes. An alternative is to use active learning based on multi-output Gaussian processes (MOGPs), which explicitly model cross-output covariance, allowing observations of one output to reduce posterior uncertainty in related outputs~\cite{zhang2016near}. When outputs can be observed separately, such methods may jointly select the sampling location and the output to be observed. However, this setting differs from that considered here, where each model evaluation returns all QoIs simultaneously and the sampling decision concerns only a common input location. Moreover, the effectiveness of MOGP-based sampling also depends on the appropriate specification of covariance kernels and reliable hyperparameters~\cite{liu2018remarks}.

Motivated by these considerations, in this work we propose a correlation-aware extension of the multi-output $\Theta$ criterion based on principal component analysis (PCA) for adaptive PCE construction. At each adaptive stage, PCA is performed on the standardized model responses at the current experimental design. It identifies orthogonal response directions according to the empirical correlation structure of the available observations. The retained directions are then used to construct the sampling criterion, with their local variance contributions normalized separately before aggregation. This combination reduces the repeated weighting of shared response patterns, while the component-retention rule limits the number of directions involved in sample selection. This methodology considers linear dependence among the outputs without requiring an explicit cross-output covariance model. Notably, PCA is used only to guide the selection of new input samples, whereas the final PCE surrogates remain defined for the original QoIs. The variance-based exploitation component of the $\Theta$ criterion is evaluated using the retained principal components, while the distance-based exploration component continues to promote coverage of the input domain, which preserves the original exploitation-exploration mechanism of the $\Theta$ criterion while accounting for correlations among multiple outputs. Moreover, an adaptive weighting based on the concentration of variance contributions is introduced to prevent the variance term from dominating sample selection at the expense of spatial coverage in problems with localized response features. Numerical examples with different output dimensions and correlation structures are used to assess its prediction accuracy, convergence behavior, and computational efficiency compared to directly aggregated multi-output $\Theta$ criterion and non-adaptive EDs.

\section{Non-intrusive polynomial chaos expansion}

Let $\mathcal{M}$ denote a computational model with random input vector $\mathbf{X}=(X_1,\ldots,X_M)$ and the generally resulting random vector response $\mathbf{Y}=\mathcal{M}(\mathbf{X})$. In UQ tasks, the objective is not only to approximate the mapping relationship between input and output, but also to characterize the statistical behavior of the model response induced by the input uncertainty. PCE is particularly attractive for this purpose because it represents the model response using polynomial bases associated with the probability distributions of the input variables. Owing to the orthonormality of the basis, statistical moments and variance-based sensitivity indices can be obtained directly from the expansion coefficients without additional evaluations of the original model \cite{sudret2008global}.

For independent input variables, the random vector $\mathbf{X}$ can be mapped to a standard random germ
$\boldsymbol{\xi}=(\xi_1,\ldots,\xi_M)$ with prescribed marginal distributions. A family of univariate orthonormal polynomials is then selected according to the distribution of each random variable. The orthonormality condition is defined as
\begin{equation}
    \langle \psi_j, \psi_k \rangle
    =
    \int_{\Omega_i}
    \psi_j(\xi_i)\psi_k(\xi_i)
    f_{\xi_i}(\xi_i)\,\mathrm{d}\xi_i
    =
    \delta_{jk},
\end{equation}
where $\psi_j$ and $\psi_k$ are the univariate polynomial basis terms associated with the corresponding random germs, respectively, $\Omega_i$ denotes the support of $\xi_i$, $f_{\xi_i}$ is its probability density function, and $\delta_{jk}$ is the Kronecker delta. The multivariate basis functions are constructed by a tensor product of these univariate bases
\begin{equation}
    \Psi_{\boldsymbol{\alpha}}(\boldsymbol{\xi})
    =
    \prod_{i=1}^{M}
    \psi_{\alpha_i}(\xi_i),
\end{equation}
where $\boldsymbol{\alpha}=(\alpha_1,\ldots,\alpha_M)$ is a multi-index specifying the polynomial degree in each germ dimension.

The vector-valued model response can then be approximated in terms of individual scalar responses as
\begin{equation}
    Y
    =
    \mathcal{M}(\mathbf{X})
    \approx
    Y^{\mathrm{PCE}}(\boldsymbol{\xi})
    =
    \sum_{\boldsymbol{\alpha}\in\mathcal{A}}
    \beta_{\boldsymbol{\alpha}}
    \Psi_{\boldsymbol{\alpha}}(\boldsymbol{\xi}),
\end{equation}
where $\beta_{\boldsymbol{\alpha}}$ are deterministic coefficients and $\mathcal{A}$ denotes a finite truncation set of the expansion. A common choice is the total-degree truncation,
\begin{equation}
    \mathcal{A}_p
    =
    \left\{
    \boldsymbol{\alpha}\in\mathbb{N}^M:
    |\boldsymbol{\alpha}|\leq p
    \right\},
\end{equation}
where $p$ is the maximum polynomial order. The number of retained basis functions is therefore
\begin{equation}
    P = |\mathcal{A}_p| = \binom{M+p}{p}.
\end{equation}
This number increases rapidly with both the input dimension and the polynomial order, which is commonly referred to as the curse of dimensionality. Consequently, the choice of $p$ and the regression strategy used to estimate the coefficients are important to balance accuracy, stability, and computational cost.

In non-intrusive PCE, the coefficients are estimated from an ED without modifying the original computational model \cite{sudret2008global, xiu2010numerical}. Collecting the scalar coefficients into the vector
$\bbeta =
(\beta_{\boldsymbol{\alpha}^{(1)}},\ldots,
 \beta_{\boldsymbol{\alpha}^{(P)}})^T$,
where $\boldsymbol{\alpha}^{(1)},\ldots,\boldsymbol{\alpha}^{(P)}$ enumerate the multi-indices in $\mathcal{A}$, the ordinary least squares (OLS) estimate is obtained from
\begin{equation}
    \bbeta
    =
    \arg\min_{\hat{\bbeta}}
    \left\|
    \mathbf{y}
    -
    \bm{\Psi}\hat{\bbeta}
    \right\|_2^2,
\end{equation}
where $\bm{\Psi}$ is the design matrix that contains the evaluations of the basis functions on the training samples. 
When the candidate polynomial space is significantly larger, sparse regression techniques such as least angle regression (LAR) can be used to select a reduced set of active basis terms \cite{blatman2011adaptive, luthen2021sparse}.

Owing to the orthonormality of the polynomial basis, the mean and variance of the surrogate response are given directly by its coefficients
\begin{equation}
    \mu_{Y^{\mathrm{PCE}}} = \beta_{\mathbf{0}},
    \qquad
    \sigma_{Y^{\mathrm{PCE}}}^2
    =
    \sum_{\boldsymbol{\alpha}\in\mathcal{A},\boldsymbol{\alpha}\neq\mathbf{0}}
    \beta_{\boldsymbol{\alpha}}^2.
\end{equation}
To relate the global output variance to individual locations in the input space, its integral representation is further considered
\begin{equation}
    \sigma_{Y^{\mathrm{PCE}}}^2
    =
    \int_{\Omega_{\boldsymbol{\xi}}}
    \sigma_{\mathcal{A}}^2(\boldsymbol{\xi})\,\dd{\boldsymbol{\xi}}
    =
    \int_{\Omega_{\boldsymbol{\xi}}}
    \underbrace{
        \left[Y^{\mathrm{PCE}}(\boldsymbol{\xi})-\beta_{\mathbf{0}}\right]^2
        f_{\boldsymbol{\xi}}(\boldsymbol{\xi})
    }_{\sigma_{\mathcal{A}}^2(\boldsymbol{\xi})}
    \,\dd{\boldsymbol{\xi}},
\end{equation}
where the integrand $ \sigma_{\mathcal{A}}^2(\boldsymbol{\xi})$ defines the local variance contribution. It can equivalently be written as
\begin{equation}
    \sigma_{\mathcal{A}}^2(\boldsymbol{\xi})
    =
    \left[
    \sum_{\boldsymbol{\alpha}\in\mathcal{A},  
    \boldsymbol{\alpha}\neq\mathbf{0}}
    \beta_{\boldsymbol{\alpha}}
    \Psi_{\boldsymbol{\alpha}}(\boldsymbol{\xi})
    \right]^2
    f_{\boldsymbol{\xi}}(\boldsymbol{\xi}).
    \label{eq:variance_contribution}
\end{equation}
This variance density describes the distribution of the overall surrogate variability over the input domain and forms the basis of the exploitation term in the adaptive $\Theta$ criterion \cite{novak2021variance}.

\section{Adaptive sequential sampling for multi-output PCE}

Adaptive sampling methods for global surrogate construction can be categorized according to the different information used to rank candidate points \cite{liu2018survey}. Geometric methods rely only on the distribution of existing samples and employ measures such as nearest-neighbor distance, minimax or maximin distance, or Voronoi-cell size to improve the coverage of the input domain \cite{johnson1990minimax,crombecq2011efficient}. By contrast, response-informed strategies use the current surrogate to guide sampling through indicators such as estimated prediction errors, predictive uncertainty, or local contributions to response variability. \cite{novak2021variance, fuhg2021state,aute2013crossvalidation}. Hybrid criteria combine both sources of information to explore insufficiently sampled regions while refining regions that are expected difficult to approximate \cite{crombecq2011novel}. Another approach is to select samples using optimality criteria regarding information matrix, such as D-, A-, and E-optimality \cite{pukelsheim2006optimal}. For global multi-output surrogates, the sampling criterion must also evaluate the collective information provided by the response vector. This differs from settings in which both the sampling location and the output to be observed can be selected, because the computational models considered here return all QoIs simultaneously at each evaluated input point.

Specifically, for regression-based PCE, adaptive strategies can improve either the regression system or the representation of the model response. Regression-oriented methods seek stable coefficient estimation through coherence-based sampling or optimality criteria derived from the PCE information matrix \cite{HADIGOL2018382}. They can be combined with basis adaptation, as in sequential $S$-optimal design and adaptive weighted least-squares PCE, the latter also supporting vector-valued responses \cite{fajraoui2017sequential,thapa2020adaptive}. For multi-output PCE, multivariate sensitivity information has also been used to construct a common anisotropic polynomial basis for the response vector \cite{loukrezis2025multivariate}. These approaches primarily adapt the polynomial basis or improve the algebraic properties of the regression system. Response-informed methods use the estimated PCE coefficients or surrogate errors directly to guide sample selection. Depending on the modeling objective, these methods emphasize different regions of the input domain. Within this class, the original $\Theta$ criterion supports global scalar-output surrogate construction by combining local variance contributions with distance-based exploration \cite{novak2021variance}. For vector-valued responses, a recent extension provides a common sampling criterion by aggregating normalized local variance contributions from individual outputs \cite{lu2026multivariate}. However, this direct output-wise aggregation does not explicitly account for cross-correlation of output. For this consideration, the following subsections first introduce this conventional formulation and then develop the proposed PCA-based extension, followed by weighting based on variance concentration.

\subsection{Conventional multi-output $\Theta$ criterion}

The local variance contribution introduced in Eq.~\eqref{eq:variance_contribution} forms the basis of the exploitation term in the adaptive $\Theta$ criterion. Consider a candidate pool $\mathcal{C}$ that contains realizations of the random vector $\boldsymbol{\xi}$. For a candidate point $\boldsymbol{\xi}^{(c)}\in\mathcal{C}$, let $\boldsymbol{\xi}^{(s)}$ denote its nearest neighbor in the current ED $\mathcal{D}_N$. The corresponding Euclidean distance is
\begin{equation}
    l_{c,s}
    =
    \left\|
    \boldsymbol{\xi}^{(c)}
    -
    \boldsymbol{\xi}^{(s)}
    \right\|_2,
    \qquad
    \boldsymbol{\xi}^{(s)}
    =
    \arg\min_{\boldsymbol{\xi}\in\mathcal{D}_N}
    \left\|
    \boldsymbol{\xi}^{(c)}-\boldsymbol{\xi}
    \right\|_2 .
    \label{eq:nearest_distance}
\end{equation}
For a scalar PCE response, the variance-based $\Theta$ criterion is defined as \cite{novak2021variance}
\begin{equation}
    \Theta\left(\boldsymbol{\xi}^{(c)}\right)
    =
    \left[
    \sigma_{\mathcal{A}}^2
    \left(\boldsymbol{\xi}^{(c)}\right)
    \sigma_{\mathcal{A}}^2
    \left(\boldsymbol{\xi}^{(s)}\right)
    \right]^{1/2}
    l_{c,s}^{M}.
    \label{eq:single_output_theta}
\end{equation}
The geometric mean of the local variance contributions provides the exploitation term, while $l_{c,s}^{M}$ promotes exploration by favoring candidates farther from their nearest neighbors in the current ED.

For a vector-valued response $\mathbf{Y}=[Y_1,\ldots,Y_{N_{\mathrm{out}}}]^T$, a PCE surrogate is constructed for each output using the same ED. The local variance contribution of the $j$th surrogate is
\begin{equation}
    \sigma_{\mathcal{A},j}^2(\boldsymbol{\xi})
    =
    \left[
    \sum_{\substack{
    \boldsymbol{\alpha}\in\mathcal{A}_j\\
    \boldsymbol{\alpha}\neq\mathbf{0}
    }}
    \beta_{\boldsymbol{\alpha}}^{(j)}
    \Psi_{\boldsymbol{\alpha}}(\boldsymbol{\xi})
    \right]^2
    f_{\boldsymbol{\xi}}(\boldsymbol{\xi}),
    \qquad
    j=1,\ldots,N_{\mathrm{out}}.
    \label{eq:output_local_variance}
\end{equation}
Because the multiple outputs may have different physical units and variance levels, their local variance contributions are normalized before aggregation. Let $\mathcal{S}$ collect the nearest neighbors corresponding to all candidates in $\mathcal{C}$. For notational convenience, define $\mathcal{Q}_c=\mathcal{C}$ and $\mathcal{Q}_s=\mathcal{S}$. The normalized aggregate of the output-wise local variance contributions is then expressed as
\begin{equation}
    v_q\left(\boldsymbol{\xi}^{(q)}\right)
    =
    \sum_{j=1}^{N_{\mathrm{out}}}
    \frac{
    \sigma_{\mathcal{A},j}^2
    \left(\boldsymbol{\xi}^{(q)}\right)
    }{
    \displaystyle
    \max_{\boldsymbol{\zeta}\in\mathcal{Q}_q}
    \sigma_{\mathcal{A},j}^2(\boldsymbol{\zeta})
    },
    \qquad
    q\in\{c,s\}.
    \label{eq:normalized_mo_variance}
\end{equation}
Finally, the conventional multi-output $\Theta$ criterion is defined by
\begin{equation}
    \Theta_{\mathrm{MO}}
    \left(\boldsymbol{\xi}^{(c)}\right)
    =
    \left[
    v_c\left(\boldsymbol{\xi}^{(c)}\right)
    v_s\left(\boldsymbol{\xi}^{(s)}\right)
    \right]^{1/2}
    l_{c,s}^{M}.
    \label{eq:multi_output_theta}
\end{equation}
The candidate maximizing Eq.~\eqref{eq:multi_output_theta} is evaluated using the original computational model and subsequently added to the ED. Since each model evaluation returns the complete response vector, the selected input sample is to update all surrogates with the aim of improving their overall predictive accuracy.

The normalization in Eq.~\eqref{eq:normalized_mo_variance} reduces the influence of differences in output scale. This aggregation counts each output separately, including outputs with similar response patterns. This treatment is consistent with assigning equal importance to all QoIs. The proposed extension instead aims to reduce the dependence of sample allocation on the repeated representation of shared response patterns. Accordingly, this conventional multi-output $\Theta$ criterion serves as the baseline for evaluating the proposed correlation-aware extension.

\subsection{PCA-based correlation-aware $\Theta$ criterion}

To consider the dependence hidden among the output components, the proposed method constructs the sampling criterion in a decorrelated principal component space. The standardized PCA is first applied to the original responses in the current ED, and the transformation is updated after each enrichment step \cite{jolliffe2016principal}.

Let $\mathbf{Y}_{\mathcal{D}}\in\mathbb{R}^{N\times N_{\mathrm{out}}}$ denote the output matrix evaluated at the current ED, where each row corresponds to one input sample and each column to one output component. The matrix is standardized by column as
\begin{equation}
\widetilde{\mathbf{Y}}_{\mathcal{D}}
=
\left(
\mathbf{Y}_{\mathcal{D}}
-
\mathbf{e}_N\boldsymbol{\mu}^{T}
\right)
\mathbf{D}_{s}^{-1},
\label{eq:output_standardization}
\end{equation}
where $\mathbf{e}_N\in\mathbb{R}^{N}$ denotes a column vector whose elements are all equal to one, $\boldsymbol{\mu}=[\mu_1,\ldots,\mu_{N_{\mathrm{out}}}]^T$ contains the sample means of the output components, and $\mathbf{D}_{s}=\operatorname{diag}(s_1,\ldots,s_{N_{\mathrm{out}}})$ contains their sample standard deviations. This standardization prevents differences in output scales from dominating the subsequent component analysis. The sample correlation matrix and its eigen-decomposition are
\begin{equation}
    \mathbf{R}
    =
    \frac{1}{N-1}
    \widetilde{\mathbf{Y}}_{\mathcal{D}}^{T}
    \widetilde{\mathbf{Y}}_{\mathcal{D}}
    =
    \mathbf{V}
    \boldsymbol{\Lambda}
    \mathbf{V}^{T},
    \label{eq:pca_decomposition}
\end{equation}
where $\boldsymbol{\Lambda} =\operatorname{diag}(\lambda_1,\ldots,\lambda_{N_{\mathrm{out}}})$ contains the eigenvalues in descending order and $\mathbf{V}=[\mathbf{v}_1,\ldots,\mathbf{v}_{N_{\mathrm{out}}}]$ contains the corresponding orthonormal eigenvectors. The standardized outputs are transformed according to
\begin{equation}
    \mathbf{Z}_{\mathcal{D}}
    =
    \widetilde{\mathbf{Y}}_{\mathcal{D}}
    \mathbf{V},
    \qquad
    \frac{1}{N-1}
    \mathbf{Z}_{\mathcal{D}}^{T}
    \mathbf{Z}_{\mathcal{D}}
    =
    \boldsymbol{\Lambda}.
    \label{eq:pca_scores}
\end{equation}
Thus, the transformed component scores have zero sample cross-covariances over the current ED, and their sample variances are given by the corresponding eigenvalues. This decorrelation refers to the empirical distribution of the available design points. It does not generally imply that the transformed random outputs are uncorrelated under the original input probability density $f_{\boldsymbol{\xi}}$.

This distinction becomes relevant during adaptive enrichment, because the sampling criterion deliberately redistributes points toward selected regions of the input domain. Consequently, the empirical correlation matrix characterizes the response variation represented by the current ED, which may differ from the correlation structure under the original input distribution. The present formulation uses this empirical structure to guide sample selection. Accordingly, the PCA eigenvalues and their relative contributions
are interpreted as measures of variation over the current ED, whereas the PCE moments and local variance contributions remain defined with respect to $f_{\boldsymbol{\xi}}$.

The fraction of the total standardized sample variance on the current ED associated with the $k$th principal component is
\begin{equation}
    r_k
    =
    \frac{\lambda_k}
    {\displaystyle\sum_{j=1}^{N_{\mathrm{out}}}\lambda_j}.
    \label{eq:pc_contribution}
\end{equation}
Components satisfying $r_k\geq r_{\min}$ are retained, and their index set is denoted by $\mathcal{K}$, with $K=|\mathcal{K}|$.
The retained fraction $\sum_{k\in\mathcal{K}}r_k$ therefore describes the standardized sample variance represented by these components on the current ED. It is not, in general, the fraction of output variance retained under the original input distribution. Furthermore, the individual-component threshold $r_{\min}$ does not prescribe a fixed cumulative retained fraction.

Since PCA is a linear transformation of the outputs, the same transformation can be applied directly to their PCE coefficients. If the surrogates contain different active polynomial terms (e.g. obtained by separate sparse solvers), their coefficient vectors are first expressed with respect to the common index set with zero coefficients assigned to absent terms
\begin{equation}
    \mathcal{A}_{\cup}
    =
    \bigcup_{j=1}^{N_{\mathrm{out}}}\mathcal{A}_j.
    \label{eq:union_basis}
\end{equation}
Let $\widetilde{\mathbf{B}}$ denote the resulting coefficient matrix of the standardized output PCEs. The coefficient matrix in the principal component space is then given by
\begin{equation}
    \mathbf{B}'
    =
    \widetilde{\mathbf{B}}\mathbf{V}.
    \label{eq:pca_coefficient_rotation}
\end{equation}

Here, coefficient standardization uses the same sample means
$\mu_j$ and sample standard deviations $s_j$ as
Eq.~\eqref{eq:output_standardization}. Specifically,
\begin{equation}
    \widetilde{\beta}_{\mathbf{0}}^{(j)}
    =
    \frac{\beta_{\mathbf{0}}^{(j)}-\mu_j}{s_j},
    \qquad
    \widetilde{\beta}_{\boldsymbol{\alpha}}^{(j)}
    =
    \frac{\beta_{\boldsymbol{\alpha}}^{(j)}}{s_j},
    \quad \boldsymbol{\alpha}\neq\mathbf{0}.
    \label{eq:standardized_pce_coefficients}
\end{equation}
The sample mean $\mu_j$ need not coincide with the original
PCE constant coefficient $\beta_{\mathbf{0}}^{(j)}$, which is
the surrogate mean under the input probability density
$f_{\boldsymbol{\xi}}$. Consequently, standardization using
the ED statistics does not generally produce a zero constant
PCE coefficient.

For a retained component $k\in\mathcal{K}$, the transformed PCE and its local variance contribution are
\begin{equation}
    Z_k^{\mathrm{PCE}}(\boldsymbol{\xi})
    =
    \sum_{\boldsymbol{\alpha}\in\mathcal{A}_{\cup}}
    \beta_{\boldsymbol{\alpha}}^{\prime(k)}
    \Psi_{\boldsymbol{\alpha}}(\boldsymbol{\xi}),
    \qquad
    \sigma_{\mathcal{A},k}^{2,\mathrm{PCA}}(\boldsymbol{\xi})
    =
    \left[
    \sum_{\substack{
    \boldsymbol{\alpha}\in\mathcal{A}_{\cup}\\
    \boldsymbol{\alpha}\neq\mathbf{0}
    }}
    \beta_{\boldsymbol{\alpha}}^{\prime(k)}
    \Psi_{\boldsymbol{\alpha}}(\boldsymbol{\xi})
    \right]^2
    f_{\boldsymbol{\xi}}(\boldsymbol{\xi}).
    \label{eq:pca_local_variance}
\end{equation}

The local contributions of the retained components are normalized and aggregated in the same manner as in Eq.~\eqref{eq:normalized_mo_variance}:
\begin{equation}
    v_q^{\mathrm{PCA}}
    \left(\boldsymbol{\xi}^{(q)}\right)
    =
    \sum_{k\in\mathcal{K}}
    \frac{
    \sigma_{\mathcal{A},k}^{2,\mathrm{PCA}}
    \left(\boldsymbol{\xi}^{(q)}\right)
    }{
    \displaystyle
    \max_{\boldsymbol{\zeta}\in\mathcal{Q}_q}
    \sigma_{\mathcal{A},k}^{2,\mathrm{PCA}}
    (\boldsymbol{\zeta})
    },
    \qquad
    q\in\{c,s\}.
    \label{eq:normalized_pca_variance}
\end{equation}

The proposed PCA-based criterion is finally defined by
\begin{equation}
    \Theta_{\mathrm{PCA}}
    \left(\boldsymbol{\xi}^{(c)}\right)
    =
    \left[
    v_c^{\mathrm{PCA}}
    \left(\boldsymbol{\xi}^{(c)}\right)
    v_s^{\mathrm{PCA}}
    \left(\boldsymbol{\xi}^{(s)}\right)
    \right]^{1/2}
    l_{c,s}^{M}.
    \label{eq:pca_theta}
\end{equation}

The effect of the proposed criterion results from the combination of \textit{rotation}, component-wise \textit{normalization}, and component \textit{selection}. 
To distinguish these operations, let
$\widetilde{Y}_j^{\mathrm{PCE}}
=(Y_j^{\mathrm{PCE}}-\mu_j)/s_j$
denote the surrogate standardized using the ED statistics.
For the variance contribution, its constant coefficient is
subsequently removed. The resulting entries of
$\mathbf{a}(\boldsymbol{\xi})$ are
\begin{equation}
    a_j(\boldsymbol{\xi})
    =
    \widetilde{Y}_j^{\mathrm{PCE}}(\boldsymbol{\xi})
    -
    \widetilde{\beta}_{\mathbf{0}}^{(j)}
    =
    \frac{
        Y_j^{\mathrm{PCE}}(\boldsymbol{\xi})
        -
        \beta_{\mathbf{0}}^{(j)}
    }{s_j},
    \qquad j=1,\ldots,N_{\mathrm{out}}.
    \label{eq:centered_standardized_pce_vector}
\end{equation}
This second centering is with respect to the original input
probability measure and follows directly from the definition
of the local variance contribution.

Let $z_k(\boldsymbol{\xi})
=Z_k^{\mathrm{PCE}}(\boldsymbol{\xi})
-\beta_{\mathbf{0}}^{\prime(k)}$
denote the transformed surrogate with its constant coefficient
removed. Since removing the constant terms commutes with the
linear output transformation,
$\mathbf{z}(\boldsymbol{\xi})
=\mathbf{V}^{T}\mathbf{a}(\boldsymbol{\xi})$.
Orthogonality of $\mathbf{V}$ then gives, when all components
are retained,
\begin{equation}
    \sum_{k=1}^{N_{\mathrm{out}}}z_k^2(\boldsymbol{\xi})
    =
    \left\|\mathbf{V}^{T}\mathbf{a}(\boldsymbol{\xi})\right\|_2^2
    =
    \left\|\mathbf{a}(\boldsymbol{\xi})\right\|_2^2
    =
    \sum_{j=1}^{N_{\mathrm{out}}}a_j^2(\boldsymbol{\xi}).
    \label{eq:pca_rotation_invariance}
\end{equation}
Multiplication by the common input density
$f_{\boldsymbol{\xi}}(\boldsymbol{\xi})$ preserves this equality.
An orthogonal rotation alone therefore does not change the
unnormalized sum of local variance contributions of the
standardized surrogates.

The sampling criterion nevertheless changes because the rotated contributions are normalized separately and only selected components are retained. Component-wise maximum normalization does not generally commute with an orthogonal rotation. Consequently, even if all numerically nonzero components are retained, rotation followed by normalization can change the relative scores of candidate points. The retention rule introduces a further change by excluding selected response directions from the criterion.

For illustration, consider an isolated block of $m$ identical standardized responses $a(\boldsymbol{\xi})$, uncorrelated with the remaining outputs under the empirical correlation structure used for PCA. Their variation is concentrated in one component, $z_1(\boldsymbol{\xi})=\sqrt{m}\,a(\boldsymbol{\xi})$. Its unnormalized local variance contribution still contains the multiplicity factor $m$. For a positive normalization denominator, however,
\begin{equation}
    \frac{
        m a^2(\boldsymbol{\xi})
        f_{\boldsymbol{\xi}}(\boldsymbol{\xi})
    }{
        \displaystyle
        \max_{\boldsymbol{\zeta}\in\mathcal{Q}_q}
        m a^2(\boldsymbol{\zeta})
        f_{\boldsymbol{\xi}}(\boldsymbol{\zeta})
    }
    =
    \frac{
        a^2(\boldsymbol{\xi})
        f_{\boldsymbol{\xi}}(\boldsymbol{\xi})
    }{
        \displaystyle
        \max_{\boldsymbol{\zeta}\in\mathcal{Q}_q}
        a^2(\boldsymbol{\zeta})
        f_{\boldsymbol{\xi}}(\boldsymbol{\zeta})
    },
    \qquad q\in\{c,s\}.
    \label{eq:pca_duplicate_normalization}
\end{equation}
The multiplicity factor thus cancels after normalization, whereas direct output-wise aggregation counts the same normalized pattern $m$ times. This example illustrates how PCA combined with component-wise normalization can reduce repeated weighting of a shared response pattern. It is not a general invariance result for arbitrary output duplication, which can also change the PCA directions and retention decisions.

Accordingly, PCA determines the response directions, normalization determines how their local contributions enter the aggregate, and truncation determines which directions participate at all.
In particular, a retained component with a small eigenvalue is not necessarily assigned a proportionally small influence: its normalized contribution can attain one, just as that of a component with a larger eigenvalue. These operations define a sampling preference among response patterns rather than a measure of the physical importance of the original QoIs.

PCA is used only to construct the sampling criterion. The original model is evaluated at the selected input points, and the PCE surrogates remain defined for the original outputs. Component truncation therefore removes directions from the sampling criterion; it does not directly truncate the final output predictions.

\subsection{Adaptive weighting of the $\Theta$ criterion based on variance concentration}
\label{sec:adaptive_weighting}

The PCA-based formulation adjusts the relative influence of shared and distinct response patterns on sample selection. However, it does not by itself prevent samples from concentrating around localized response features. In both the conventional and PCA-based $\Theta$ criteria, the variance-based and distance-based terms are combined through direct multiplication. Although the distance term favors poorly sampled regions, a large distance may not compensate for a very small local variance contribution. Consequently, candidates near pronounced response features may continue to dominate the selection, even when other parts of the domain remain sparsely sampled. PCA can redistribute samples among different features without necessarily improving coverage away from them.

This distinction matters for \emph{global polynomial approximation}. Accurate representation of localized features is important, but the stability and predictive accuracy of a high-degree polynomial fit also depend on the distribution of samples throughout the domain, including regions near its boundaries. To moderate excessive concentration while preserving the response-informed character of the criterion, we introduce an adaptive exponent for the variance-based term.

Let $N_c$ denote the number of available candidate points, excluding those already included in the ED, and define
\[
V_i
=
\left[
v_c\left(\boldsymbol{\xi}^{(c_i)}\right)
v_s\left(\boldsymbol{\xi}^{(s_i)}\right)
\right]^{1/2},
\qquad i=1,\ldots,N_c.
\]
Here, $v_c$ and $v_s$ are the normalized aggregates for either the original outputs or the retained principal components, depending on the formulation used. The adaptively weighted criterion is
\begin{equation}
\Theta_{\eta}\left(\boldsymbol{\xi}^{(c_i)}\right)
=
V_i^{\eta}l_{c_i,s_i}^{M},
\label{eq:adaptive_weighted_theta}
\end{equation}
where the exponent $\eta$ is common to all candidates at a given selection step. For positive scores and distances, taking the logarithm gives
\begin{equation}
\log\Theta_{\eta}\left(\boldsymbol{\xi}^{(c_i)}\right)
=
\eta\log V_i
+
M\log l_{c_i,s_i}.
\label{eq:adaptive_weighted_theta_log}
\end{equation}
Thus, choosing $0<\eta<1$ reduces the contrast between positive variance-based scores and gives distance differences greater relative influence on sample selection. The ordering induced by the variance-based term alone is preserved, but its influence relative to exploration is moderated. For example, a ratio of $100$ between two positive variance scores becomes
$100^{0.2}\approx2.51$ when $\eta=0.2$.

The strength of this adjustment is determined from the concentration of the current variance-based scores:
\begin{equation}
\eta
=
\frac{
\left(\displaystyle\sum_{i=1}^{N_c}V_i\right)^2
}{
N_c\displaystyle\sum_{i=1}^{N_c}V_i^2
}.
\label{eq:variance_concentration_exponent}
\end{equation}
For nonnegative scores with at least one positive value, $1/N_c\leq\eta\leq1$. Equal positive scores give $\eta=1$, whereas concentration of the scores among fewer candidates produces a smaller exponent.

This dependence can be expressed directly in terms of relative variability. Define the mean score 
    $\overline{V}=N_c^{-1}\sum_{i=1}^{N_c}V_i$ 
and its variance across the candidate pool
    $s_V^2=N_c^{-1}\sum_{i=1}^{N_c}(V_i-\overline{V})^2$.
For $\overline{V}>0$, the coefficient of variation is
$\mathrm{CoV}_V=s_V/\overline{V}$, and
Eq.~\eqref{eq:variance_concentration_exponent} becomes
\begin{equation}
\eta
=
\frac{\overline{V}^{\,2}}
{\overline{V}^{\,2}+s_V^2}
=
\frac{1}{1+\mathrm{CoV}_V^2}.
\label{eq:variance_concentration_cv}
\end{equation}
The exponent therefore approaches one when the scores have low relative variability and decreases as their relative variability increases. Highly concentrated scores consequently trigger stronger compression of the variance-based contrast. As an illustrative limiting case, if exactly $k$ candidates have the same positive score and all remaining candidates have zero scores, then $\eta=k/N_c$.
The exponent describes concentration over the available candidate pool; it should not generally be interpreted as a fraction of the physical domain.

When all scores are equal and positive, the variance-based factor is constant across candidates, and maximizing Eq.~\eqref{eq:adaptive_weighted_theta} is equivalent to sequential maximin selection. More generally, the adaptive weighting moderates the influence of concentrated variance contributions, but it does not guarantee coverage of every region. If all variance-based scores vanish, the exponent is not evaluated and the next point is selected by maximizing its distance to the current ED.

\section{Numerical Examples}
\label{sec:numerical_examples}

The proposed PCA-based sampling strategy is evaluated through the following multi-output numerical examples with different response characteristics. To ensure a consistent comparison of the sampling strategies and avoid possible effects introduced by sparse regression, all PCE coefficients are estimated using OLS. Six sampling strategies are compared in the two-dimensional examples: random MC sampling (Random-OLS), LHS (LHS-OLS), the conventional multi-output $\Theta$ criterion ($\Theta$-OLS), the proposed PCA-based criterion (PCA-$\Theta$-OLS), and their adaptively weighted versions ($\eta$-$\Theta$-OLS and $\eta$-PCA-$\Theta$-OLS). These comparisons examine the effects of PCA-based aggregation and adaptive weighting separately and in combination.

To evaluate the statistical performance of the sampling strategies, the two-dimensional examples are repeated 50 times. Within each repetition, all methods share the same initial ED. Random-OLS adds independent random samples, whereas LHS-OLS generates an independent Latin hypercube design for each subsequent training budget. For adaptive sampling, additional points are selected sequentially within each batch using fixed PCE surrogates and a fixed PCA transformation. Within each batch, every selected point is temporarily added to the ED, and the nearest neighbors, criterion normalization factors, and adaptive exponent are updated before selecting the next point. Once the entire batch has been selected, the original model is evaluated at the selected points and the surrogates are refitted using the enriched ED. The distance-based term is evaluated in the standardized input space. Since the inputs considered in this study are uniform over their respective intervals $[a_m,b_m]$, the transformation $\xi_m=2(x_m-a_m)/(b_m-a_m)-1$ maps them to uniform distributions on $[-1,1]$.

Two different strategies are considered for adaptive selection of the polynomial degree. In the common-degree setting, all output surrogates share a single degree selected by minimizing the maximum normalized leave-one-out (LOO) error across the outputs:
\begin{equation}
p_{\mathrm{com}}^{*}
=
\underset{p\in\mathcal{P}_N}{\arg\min}
\max_{j=1,\ldots,N_{\mathrm{out}}}
\varepsilon_{\mathrm{LOO},j}(p).
\label{eq:common_order_selection}
\end{equation}
This minimax rule selects the degree according to the largest output-wise LOO error. Alternatively, in the individual-degree setting, the polynomial degree is selected independently for each output:
\begin{equation}
p_j^{*}
=
\underset{p\in\mathcal{P}_N}{\arg\min}
\varepsilon_{\mathrm{LOO},j}(p),
\qquad
j=1,\ldots,N_{\mathrm{out}},
\label{eq:individual_order_selection}
\end{equation}
where $\mathcal{P}_N$ denotes the admissible candidate degrees at training size $N$. This setting adapts the polynomial degree to the approximation characteristics of each output, although the resulting surrogates may have different model complexities. For OLS-based PCE, LOO errors can be evaluated analytically from the current regression system. Each degree-selection strategy therefore uses its respective current ED without requiring additional evaluations of the original computational model \cite{blatman2011adaptive}. The output-wise LOO error is normalized by the empirical training variance:
\begin{equation}
\varepsilon_{\mathrm{LOO},j}(p)
=
\frac{1}{N\widehat{v}_j^{\mathrm{train}}}
\sum_{i=1}^{N}
\left[
\frac{
y_j^{(i)}-\widehat{Y}_{j,p}\left(\boldsymbol{\xi}^{(i)}\right)
}{
1-h_{ii}(p)
}
\right]^2,
\qquad
\widehat{v}_j^{\mathrm{train}}
=
\frac{1}{N-1}
\sum_{i=1}^{N}
\left(y_j^{(i)}-\overline{y}_j^{\mathrm{train}}\right)^2,
\label{eq:normalized_loo}
\end{equation}
where $\overline{y}_j^{\mathrm{train}}$ denotes the training-sample mean and $h_{ii}(p)$ is the corresponding diagonal entry of the hat matrix. To improve the numerical stability of the LOO calculation, we use the singular value decomposition (SVD) $\bm{\Psi}_p=\mathbf{U}_p\mathbf{S}_p\mathbf{W}_p^{\mathsf T}$ and compute the leverage values directly from the left singular vectors as $h_{ii}(p)=\sum_{k=1}^{P}(\mathbf{U}_p)_{ik}^{2}$, avoiding explicit formation of the Gram matrix. Only candidates with $N>P$ are considered, where $P$ is the number of basis terms. Two additional checks are applied: polynomial candidates are rejected if $\sigma_{\min}\leq\epsilon_{\mathrm{rank}}\sigma_{\max}$, indicating numerical rank deficiency, or if any LOO denominator satisfies $1-h_{ii}(p)\leq\epsilon_{\mathrm{LOO}}$. Here, $\sigma_{\min}$ and $\sigma_{\max}$ are the smallest and largest singular values in $\mathbf{S}_p$, while $\epsilon_{\mathrm{rank}}$ and $\epsilon_{\mathrm{LOO}}$ are numerical tolerances determined by machine precision and matrix size. 

Surrogate accuracy is evaluated on an independent test set using the root mean square error (RMSE), mean absolute error (MAE), and maximum absolute error (MaxAE). For the $j$th output, these measures are defined as
\begin{equation}
\mathrm{RMSE}_j
=
\left[
\frac{1}{N_{\mathrm{test}}}
\sum_{i=1}^{N_{\mathrm{test}}}
\left(e_{ij}\right)^2
\right]^{1/2},
\qquad
\mathrm{MAE}_j
=
\frac{1}{N_{\mathrm{test}}}
\sum_{i=1}^{N_{\mathrm{test}}}
\left|e_{ij}\right|,
\qquad
\mathrm{MaxAE}_j
=
\max_{i=1,\ldots,N_{\mathrm{test}}}
\left|e_{ij}\right|,
\label{eq:output_errors}
\end{equation}
where
\begin{equation}
e_{ij}
=
y_{j,\mathrm{test}}^{(i)}
-
\widehat{Y}_j
\left(\boldsymbol{\xi}_{\mathrm{test}}^{(i)}\right)
\end{equation}
is the prediction error at the $i$th test point. Test points are generated independently and are not used for degree selection or adaptive sampling. The same test set is used across methods, training budgets, and repetitions. For the random field (RF) example, the field realization changes between repetitions but is shared by all methods within each repetition, and the reference responses on the test set are evaluated for the corresponding realization. To account for differences in output scale, all three error measures are normalized using the root-mean-square magnitude of the corresponding reference response:
\begin{equation}
\begin{gathered}
\mathrm{NRMSE}_j
=
\frac{\mathrm{RMSE}_j}{s_j^{\mathrm{test}}},
\qquad
\mathrm{NMAE}_j
=
\frac{\mathrm{MAE}_j}{s_j^{\mathrm{test}}},
\qquad
\mathrm{NMaxAE}_j
=
\frac{\mathrm{MaxAE}_j}{s_j^{\mathrm{test}}},
\\[6pt]
s_j^{\mathrm{test}}
=
\left[
\frac{1}{N_{\mathrm{test}}}
\sum_{i=1}^{N_{\mathrm{test}}}
\left(y_{j,\mathrm{test}}^{(i)}\right)^2
\right]^{1/2}.
\end{gathered}
\label{eq:normalized_output_errors}
\end{equation}
The aggregate NRMSE and NMAE are obtained by averaging their output-wise values, whereas the aggregate NMaxAE is defined as the largest NMaxAE among all outputs. NRMSE emphasizes relatively large prediction errors and is consistent with the least-squares construction of the PCE, while NMAE describes the average absolute prediction accuracy. NMaxAE measures the largest normalized absolute prediction error observed across all outputs and test points. For clarity, the error measures are expressed as $\log_{10}(\mathrm{error})$, with the mean and $\pm$ one standard deviation over 50 repetitions shown as curves and corresponding bands, respectively.

\subsection{Two-dimensional Gaussian fields with trigonometric perturbations}
\label{sec:gaussian_example}

The first example is designed to examine how the sampling strategies respond to localized nonlinear features that are either shared among several outputs or specific to individual outputs. The two-dimensional input domain also allows the resulting sampling patterns to be visualized directly. For this purpose, six outputs are constructed from localized Gaussian functions and trigonometric perturbations. The inputs $X_1$ and $X_2$ are independent and uniformly distributed over $[0,1]$. For simplification, a Gaussian function is denoted by
\begin{equation}
    G(\boldsymbol{x};c_1,c_2,s_1,s_2)
    =
    \exp\left[
        -\left(\frac{x_1-c_1}{s_1}\right)^2
        -\left(\frac{x_2-c_2}{s_2}\right)^2
    \right],
\end{equation}
where $(c_1,c_2)$ specifies the location of the peak and $s_1, s_2$ control its spatial distribution. With the baseline $G_0(\boldsymbol{x})=G(\boldsymbol{x};0.25,0.25,0.16,0.16)$, the outputs are as follow:
\begin{equation}
    \begin{aligned}
        Y_1(\boldsymbol{x}) &= G_0(\boldsymbol{x}),\\
        Y_2(\boldsymbol{x}) &= 1.03G_0(\boldsymbol{x})+0.02\sin(2\pi x_1),\\
        Y_3(\boldsymbol{x}) &= 0.97G_0(\boldsymbol{x})+0.03\cos(2\pi x_2),\\
        Y_4(\boldsymbol{x}) &= G_0(\boldsymbol{x})+0.03\sin\bigl(2\pi(x_1+x_2)\bigr),\\
        Y_5(\boldsymbol{x}) &= 0.85G(\boldsymbol{x};0.75,0.70,0.13,0.13),\\
        Y_6(\boldsymbol{x}) &= 0.70G(\boldsymbol{x};0.80,0.25,0.14,0.14).
    \end{aligned}
\end{equation}
The first four outputs share the same Gaussian peak, while the perturbations introduce smaller specific variations. The remaining two outputs contain peaks at different locations. This construction allows the sampling strategies to be examined when repeated response information coexists with distinct localized features.

The initial ED contains 30 uniformly distributed samples. Training budgets increase in steps of 10 to 1000 samples. The comparison is repeated 50 times using independently generated initial EDs. Within each repetition, all six methods share the same initial ED to enable a reasonable comparison. The adaptive methods use a fixed $100\times100$ candidate grid. Prediction accuracy is evaluated at 1000 independent test points, and the PCA contribution threshold is set to $r_{\min}=0.01$.

\begin{figure}[htbp!]
\centering
\includegraphics[width=0.9\linewidth]{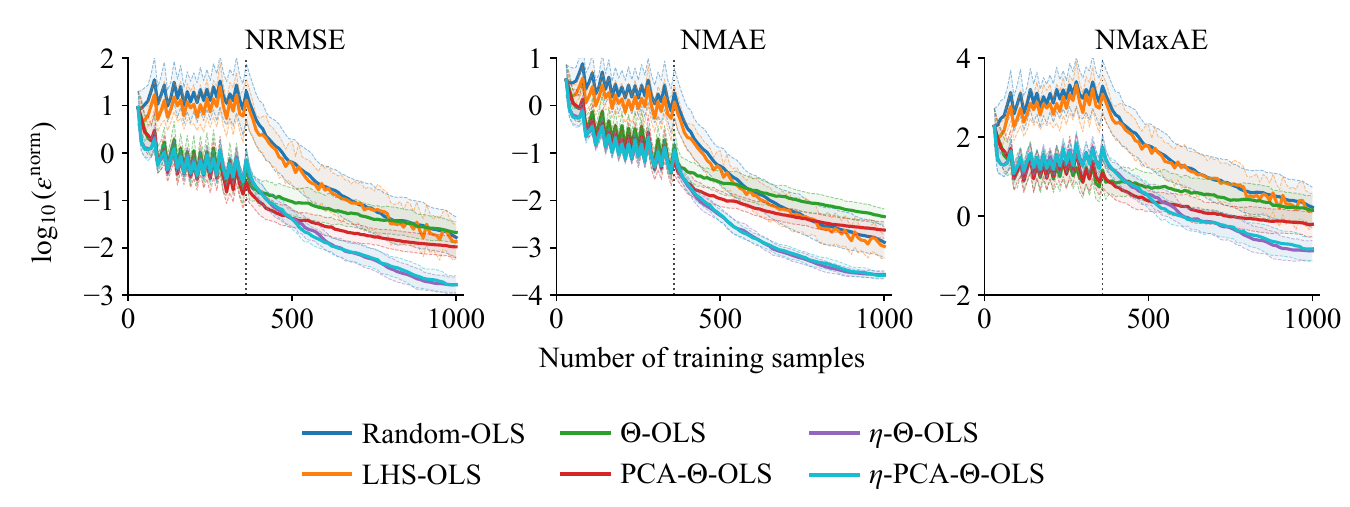}
\caption{2D Gaussian: convergence of the normalized errors with the same maximum admissible degree for all six methods (mean $\pm$ one standard deviation in logarithmic space). The vertical line marks the first training budget with $p=25$.}
\label{fig:gaussian_shared_error}
\end{figure}

\begin{figure}[!t]
    \centering
    \includegraphics[width=0.75\linewidth,height=.75\textheight,keepaspectratio]{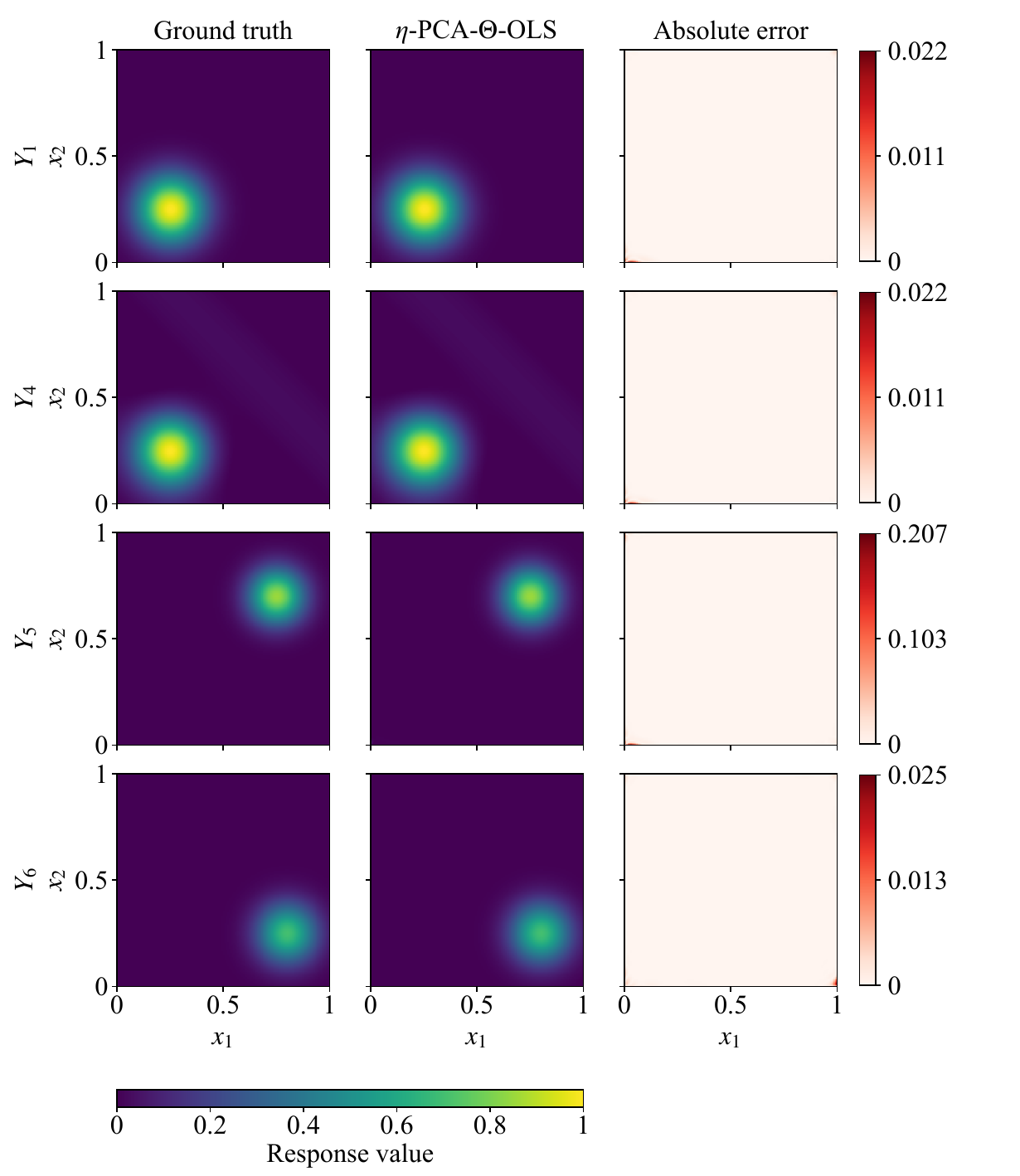}
    \caption{2D Gaussian: ground-truth fields, $\eta$-PCA-$\Theta$-OLS predictions, and absolute errors for representative outputs in the worst-prediction repeat at $N=1000$.}
    \label{fig:gaussian_field}
\end{figure}

To remove the effect of errors by different polynomial degree, all six methods first use the same polynomial degree at each training size, namely the largest $p\in\mathcal{P}=\{5,\ldots,25\}$ for which $P(p)=(p+1)(p+2)/2<N$. The three error measures exhibit pronounced fluctuations at small training sizes and converge more smoothly beyond approximately $N=360$ (Fig.~\ref{fig:gaussian_shared_error}). In two dimensions, the total-degree basis with $p=25$ contains $P=351$ terms, and, with samples added in increments of 10, the corresponding regression system first becomes overdetermined at $N=360$. Before this point, increasing the degree repeatedly introduces additional coefficients while the size of ED remains close to the basis size. Despite the early fluctuations, all four adaptive methods generally attain lower errors than Random-OLS and LHS-OLS at the same small training budget across all three measures. Once the polynomial degree reaches its upper bound, NMaxAE decreases faster for PCA-$\Theta$-OLS than for conventional $\Theta$-OLS. The benefit of adaptive weighting is most evident later in enrichment, when NRMSE, NMAE, and NMaxAE decrease more rapidly for both weighted methods than for their unweighted counterparts. By the final training budget, $\eta$-$\Theta$-OLS and $\eta$-PCA-$\Theta$-OLS give similar results and rank first and second across all three error measures. The detailed accuracy of the resulting surrogates is illustrated in Fig.~\ref{fig:gaussian_field} for $\eta$-PCA-$\Theta$-OLS at the final budget. The comparison shows representative outputs from the repetition with the largest NMaxAE for this method. Even in this repetition, the surrogates reproduce both the peak shared by $Y_1$ and $Y_4$ and the distinct peaks in $Y_5$ and $Y_6$, and the largest absolute error occurs at the boundary corner for $Y_5$.


\begin{figure}[!b]
    \centering
    \includegraphics[width=0.72\linewidth]{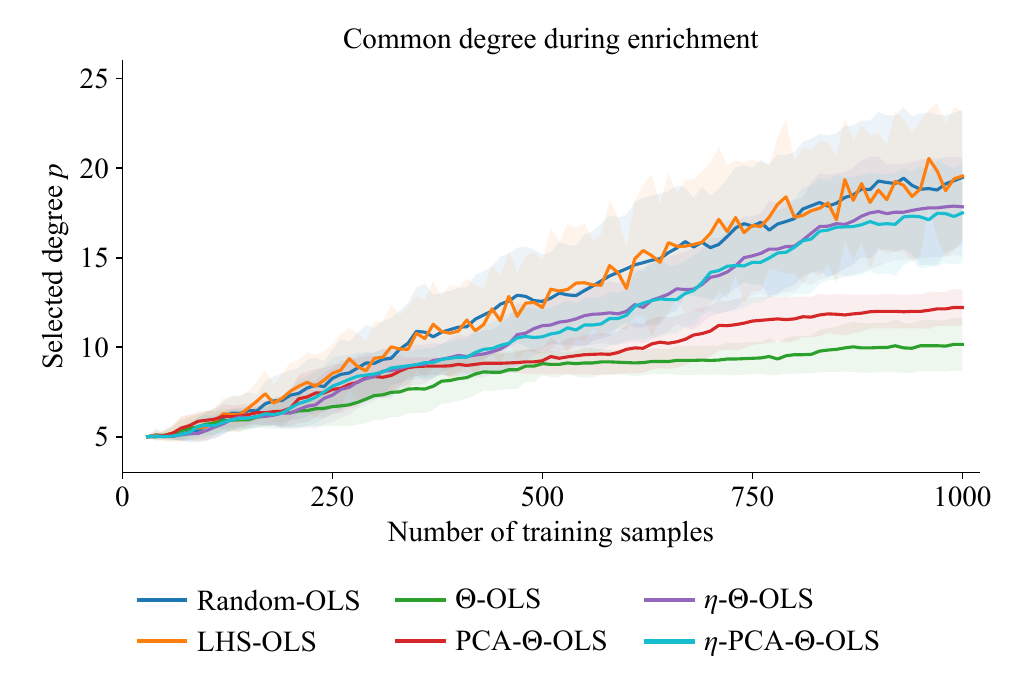}
    \caption{2D Gaussian: evolution of the common polynomial degree selected by LOO over 50 repetitions.}
    \label{fig:gaussian_p}
\end{figure}

\begin{figure}[htbp!]
    \centering
    \includegraphics[width=0.95\linewidth]{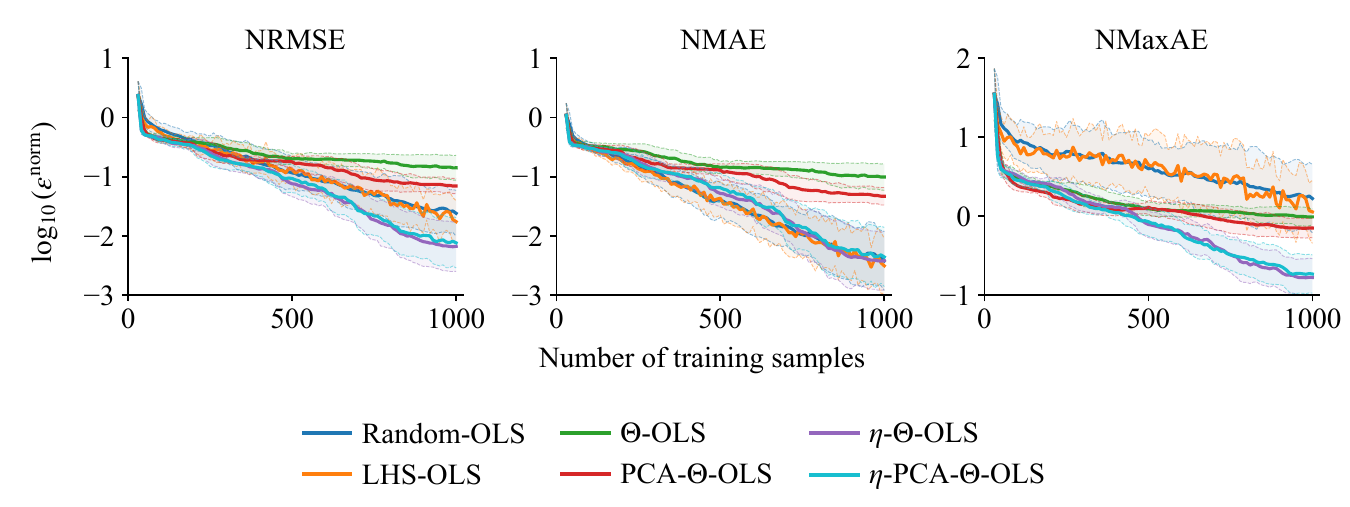}
    \caption{2D Gaussian: convergence of the normalized prediction errors with common polynomial degree selected by LOO error (mean $\pm$ one standard deviation in logarithmic space).}
    \label{fig:gaussian_convergence}
\end{figure}

Besides prediction accuracy, the stability of the surrogate is also an important consideration when selecting the polynomial degree. The LOO error reflects how accurately the surrogate predicts each training response when the corresponding point is omitted. It can be evaluated from the current ED without prior knowledge of the appropriate degree or additional model evaluations. So, we repeat the comparison using the common-degree LOO criterion introduced in Section~\ref{sec:numerical_examples}. At each training budget, each sampling method selects a single degree for all six outputs by minimizing the largest output-wise LOO error on its own ED. The selected degrees initially increase for all six methods, but their evolution differs as the ED grows (Fig.~\ref{fig:gaussian_p}). By the final budget, Random-OLS and LHS-OLS select degrees close to $20$ on average, compared with approximately $10$ for conventional $\Theta$-OLS and $12$ for PCA-$\Theta$-OLS. With adaptive weighting, the average final degree is close to $18$ for both $\Theta$ criteria. The lower degrees selected by the two unweighted $\Theta$ methods can be explained by examining the same $p=20$ fit on each final ED. For each method and repetition, we consider the output with the largest normalized LOO error. For both $\Theta$-OLS and PCA-$\Theta$-OLS, a single training point near a corner accounts for more than $80\%$ of the sum of squared LOO residuals in the median case. These points have leverage values close to one, so dividing the fitted residual by $1-h_{ii}$ substantially increases its contribution to the LOO error. Thus, a high-degree fit can receive a large LOO score because its prediction near one corner is particularly sensitive to the omission of a single sample, which further motivates the use of adaptive weighting to improve spatial coverage away from the localized feature. With the common degree selected by LOO, both $\Theta$-OLS and PCA-$\Theta$-OLS achieve a lower final NMaxAE than Random-OLS and LHS-OLS, but their final NRMSE and NMAE are higher (Fig.~\ref{fig:gaussian_convergence}). Adaptive weighting reduces all three final errors for both $\Theta$ criteria. Finally, these two weighted methods give the lowest NRMSE, NMaxAE, and NMAE.

\begin{figure}[!t]
    \centering
    \includegraphics[width=0.95\linewidth,height=.75\textheight,keepaspectratio]{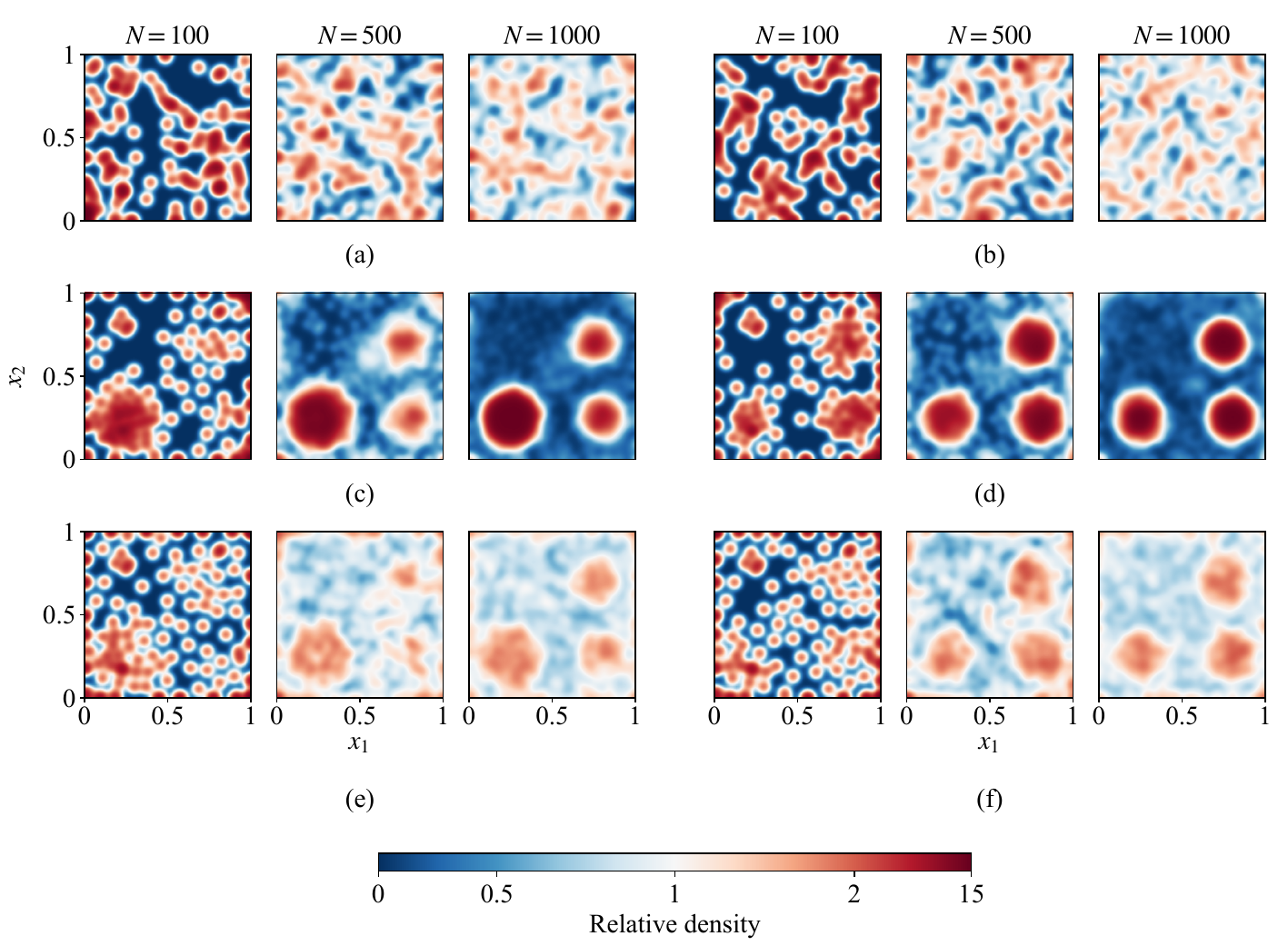}
    \caption{2D Gaussian: sampling densities at $N=100$, $500$, and $1000$ for (a) Random-OLS, (b) LHS-OLS, (c) $\Theta$-OLS, (d) PCA-$\Theta$-OLS, (e) $\eta$-$\Theta$-OLS, and (f) $\eta$-PCA-$\Theta$-OLS. A relative density of one corresponds to uniform sampling.}
    \label{fig:gaussian_sampling}
\end{figure}

The sampling densities in Fig.~\ref{fig:gaussian_sampling} explain these differences in prediction accuracy. As the training budget increases, Random-OLS and LHS-OLS cover the input domain relatively uniformly, whereas the four adaptive methods place more samples near the Gaussian peaks. Conventional $\Theta$-OLS concentrates most strongly around the peak shared by $Y_1$ to $Y_4$, while PCA-$\Theta$-OLS allocates samples more evenly among the three peak regions. With adaptive weighting, the peaks remain visible as regions of higher sampling density, but both criteria also achieve more uniform coverage of the surrounding domain. 

\begin{figure}[htbp!]
    \centering
    \includegraphics[width=0.78\linewidth]{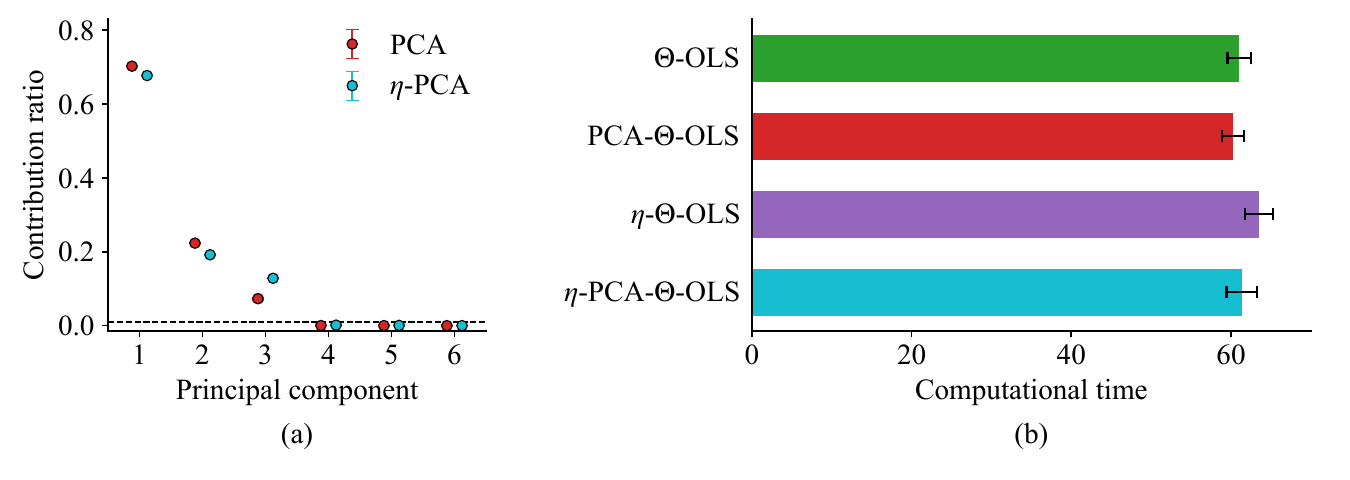}
    \caption{2D Gaussian: (a) PCA contribution ratios on the final ED, with the dashed line marking the retention threshold $r_{\min}=0.01$, and (b) mean and standard deviation of computational time for four $\Theta$ sampling methods over all training budgets and repetitions.}
    \label{fig:gaussian_pca}
\end{figure}

At the final training budget, both PCA-based methods retain three of the six principal components in all 50 repetitions (Fig.~\ref{fig:gaussian_pca}a). For PCA-$\Theta$-OLS, these components account on average for $70.3\%$, $22.4\%$, and $7.3\%$ of the total standardized sample variance on the final ED, respectively. Evaluating the sampling criterion on three retained components yields a modest saving in runtime due to the size of this example (Fig.~\ref{fig:gaussian_pca}b). With PCA, the mean runtime decreases from $61.0$ to $60.3$~s for the conventional criterion and from $63.6$ to $61.3$~s with adaptive weighting. 

Moreover, Table~\ref{tab:gaussian_peak_local_mae_fixed_p} examines how the redistribution of training samples by PCA affects local accuracy by evaluating the surrogates on dense grids around the three Gaussian peaks. For the conventional $\Theta$ criterion, PCA reduces the local MAE near the peak of $Y_5$, while the errors at the shared peak and the peak of $Y_6$ increase. With adaptive weighting, the local MAE at the shared peak increases by less than $0.3\times10^{-6}$, whereas the errors near the peaks of $Y_5$ and $Y_6$ decrease by approximately $5.7\times10^{-6}$ and $0.9\times10^{-6}$, respectively. In absolute terms, this trade-off is favourable if the three peak regions are given equal importance. All reported mean local errors are below $10^{-4}$, but the acceptability of the increase at the shared peak ultimately depends on the accuracy required for the corresponding QoIs in the engineering application.

\begin{table}[htbp!]
    \centering
    \caption{2D Gaussian: local MAE at $N=1000$ for four adaptive sampling methods using the same polynomial degree schedule and final degree $p=25$. Entries are in units of $10^{-6}$ and show the mean $\pm$ one standard deviation over 50 repetitions. Each peak region is a disk of radius $0.1$ centred at the location shown. For the shared peak, the local MAEs of $Y_1$ to $Y_4$ are averaged within each repetition.}
    \label{tab:gaussian_peak_local_mae_fixed_p}
    \begin{tabular}{@{}lccc@{}}
        \hline
        Sampling method & Shared peak ($Y_1$ to $Y_4$) & Peak of $Y_5$ & Peak of $Y_6$ \\
                        & $(0.25,0.25)$ & $(0.75,0.70)$ & $(0.80,0.25)$ \\
        \hline
        $\Theta$-OLS            & $0.94 \pm 0.39$ & $29.45 \pm 7.48$ & $3.74 \pm 1.50$  \\
        PCA-$\Theta$-OLS        & $1.84 \pm 0.55$ & $23.45 \pm 4.98$ & $4.69 \pm 2.00$  \\
        $\eta$-$\Theta$-OLS     & $5.72 \pm 0.62$ & $93.28 \pm 6.08$ & $18.99 \pm 2.09$ \\
        $\eta$-PCA-$\Theta$-OLS & $5.95 \pm 0.59$ & $87.54 \pm 6.21$ & $18.10 \pm 2.04$ \\
        \hline
    \end{tabular}
\end{table}

\subsection{One-dimensional temporal and spatial wave equation with multiple response quantities}
\label{sec:wave_example}

\begin{figure}[!h]
    \centering
    \includegraphics[width=.85\linewidth]{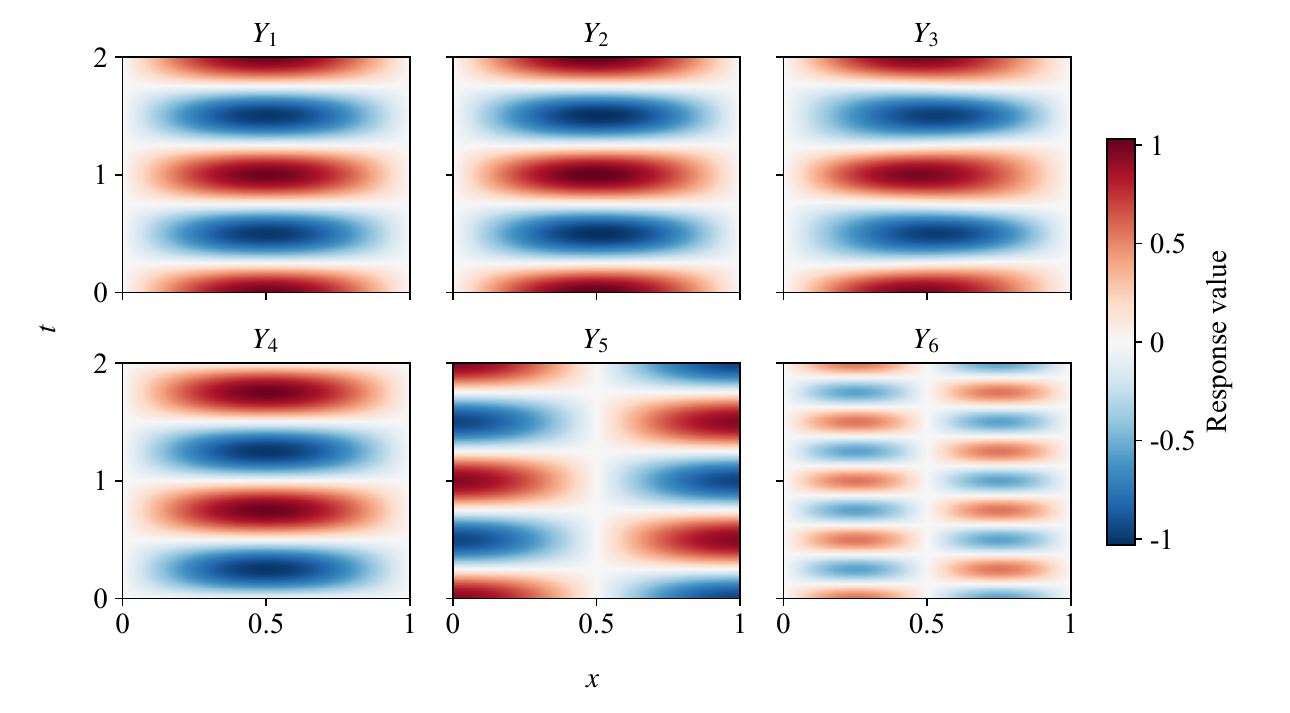}
    \caption{1D wave equation: analytical response of the six outputs over the spatial-temporal domain.}
    \label{fig:wave_outputs}
\end{figure}

The second example is designed to examine adaptive PCE construction when correlated outputs exhibit different approximation complexities. Its oscillatory and non-polynomial response features require relatively high polynomial degrees, while the different modal compositions of the outputs allow common and individual degree selection to be compared. For this purpose, consider the following one-dimensional wave equation:
\begin{equation}
    \frac{\partial^2 u}{\partial t^2}
    =
    c^2\frac{\partial^2 u}{\partial x^2},
    \qquad
    (x,t)\in[0,1]\times[0,2],
\end{equation}
with wave speed $c=2$ and homogeneous Dirichlet boundary conditions $u(0,t)=u(1,t)=0$. For the initial conditions
\begin{equation}
    u(x,0)=a\sin(\pi x)+b\sin(2\pi x),
    \qquad
    \frac{\partial u}{\partial t}(x,0)=0,
\end{equation}
the corresponding analytical solution is $u(x,t)=a u_1(x,t)+b u_2(x,t)$, where
\begin{equation}
    u_1(x,t)=\sin(\pi x)\cos(2\pi t),
    \qquad
    u_2(x,t)=\sin(2\pi x)\cos(4\pi t).
\end{equation}
The coefficients $a$ and $b$ determine the amplitudes of different wave modes, where the secondary mode $u_2$ has twice the spatial wavenumber and temporal frequency of the primary mode $u_1$. Six outputs are constructed from the two modal components, their linear combinations, and derivatives of the primary mode:
\begin{equation}
    \begin{aligned}
        Y_1(x,t) &= u_1(x,t),\\
        Y_2(x,t) &= 1.03u_1(x,t)+0.015u_2(x,t),\\
        Y_3(x,t) &= 0.97u_1(x,t)+0.05u_2(x,t),\\
        Y_4(x,t) &= 0.16\frac{\partial u_1}{\partial t}
        =-0.32\pi\sin(\pi x)\sin(2\pi t),\\
        Y_5(x,t) &= 0.30\frac{\partial u_1}{\partial x}
        =0.30\pi\cos(\pi x)\cos(2\pi t),\\
        Y_6(x,t) &= 0.55u_2(x,t),
    \end{aligned}
\end{equation}
where the first three outputs are dominated by the primary mode. The secondary mode contributes only a small fraction of the variance of $Y_2$ and $Y_3$, but its higher frequency can still affect the polynomial degree required for accurate approximation. It is also represented separately by $Y_6$. The outputs $Y_4$ and $Y_5$ are scaled temporal and spatial derivatives of the primary mode, respectively. Although $Y_5$ has the same spatial and temporal frequencies as $Y_1$, its cosine form on $x$ gives it a different spatial profile and allows nonzero values at $x=0$ and $x=1$ (Fig.~\ref{fig:wave_outputs}).

The spatial coordinate and time are modelled as independent inputs, with $X\sim\mathcal{U}(0,1)$ and $T\sim\mathcal{U}(0,2)$. As in the Gaussian example, the six sampling methods are compared over 50 repetitions with training budgets from $N=30$ to $N=1000$ in increments of 10. The adaptive methods use a $100\times100$ candidate grid and the PCA contribution threshold is $r_{\min}=0.01$. All methods are evaluated on the same independent test set of 1000 points. We compare common and individual polynomial degree selection using the LOO rules. 

\begin{figure}[htbp!]
    \centering
    \includegraphics[width=0.72\linewidth]{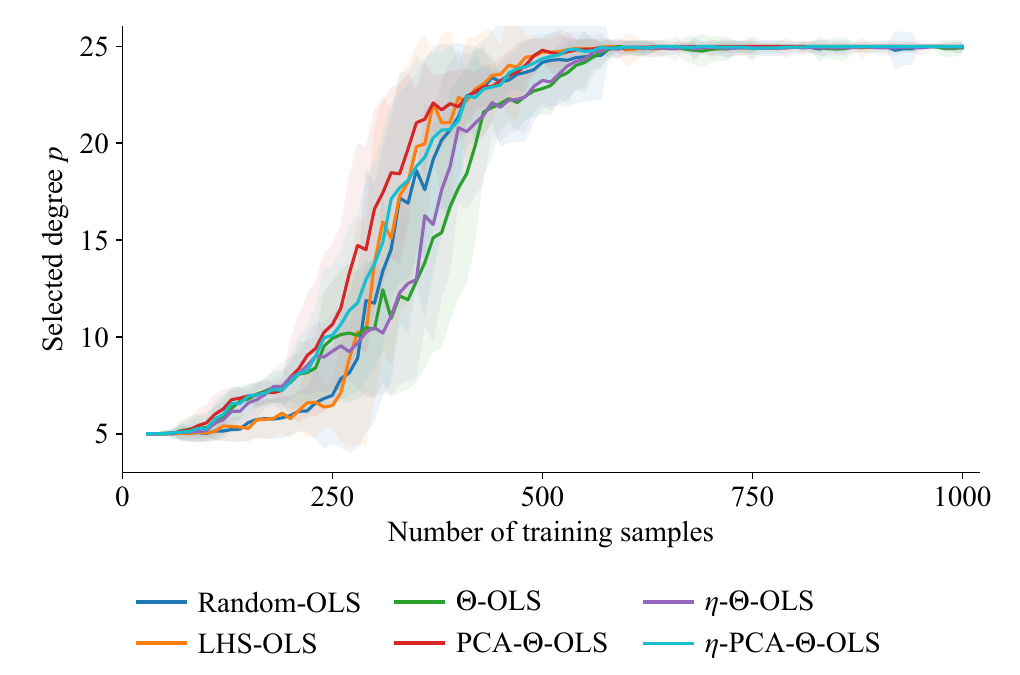}
    \caption{1D wave equation: evolution of the common polynomial degree selected by LOO over 50 repetitions.}
    \label{fig:wave_common_p}
\end{figure}

\begin{figure}[htbp!]
    \centering
    \includegraphics[width=.96\linewidth]{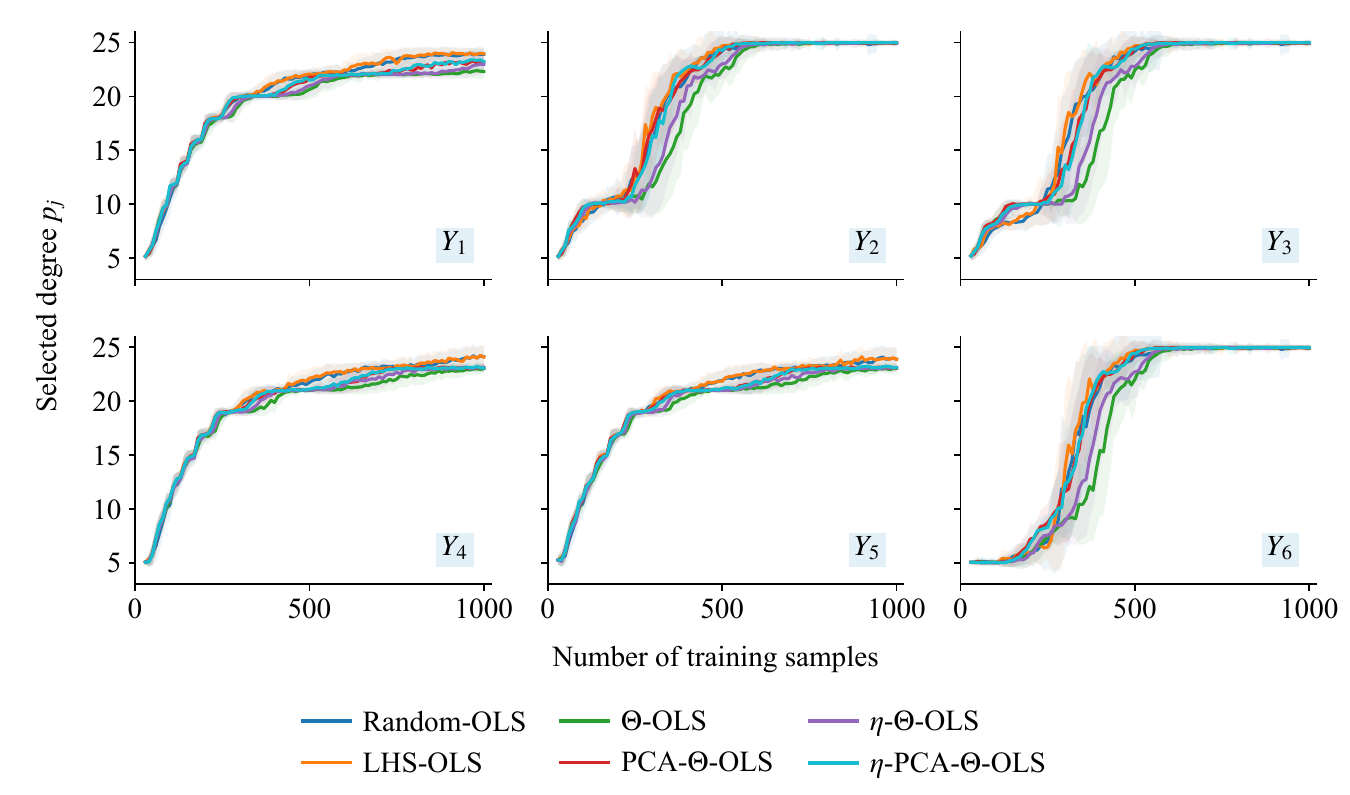}
    \caption{1D wave equation: evolution of the individual polynomial degrees selected by LOO for the six outputs over 50 repetitions.}
    \label{fig:wave_individual_p}
\end{figure}

Under common-degree selection, all six methods ultimately select a degree close to the upper bound of $p=25$, although they reach it at different training budgets (Fig.~\ref{fig:wave_common_p}). The mean selected degrees range from approximately $14$ to $21$ at $N=360$ and reach $25$ for all methods by $N=600$. Individual-degree selection reveals which outputs drive the choice of a large common basis (Fig.~\ref{fig:wave_individual_p}). At $N=1000$, the mean degrees for $Y_2$, $Y_3$, and $Y_6$ are close to $25$ for every method. The corresponding degrees for $Y_1$, $Y_4$, and $Y_5$ range from $21$ to $24$. Thus, even the weak secondary-mode contribution in $Y_2$ is associated with a higher selected degree, whereas the three primary-mode outputs admit a modest reduction in basis size.

\begin{figure}[htbp!]
    \centering
    \includegraphics[width=.91\linewidth]{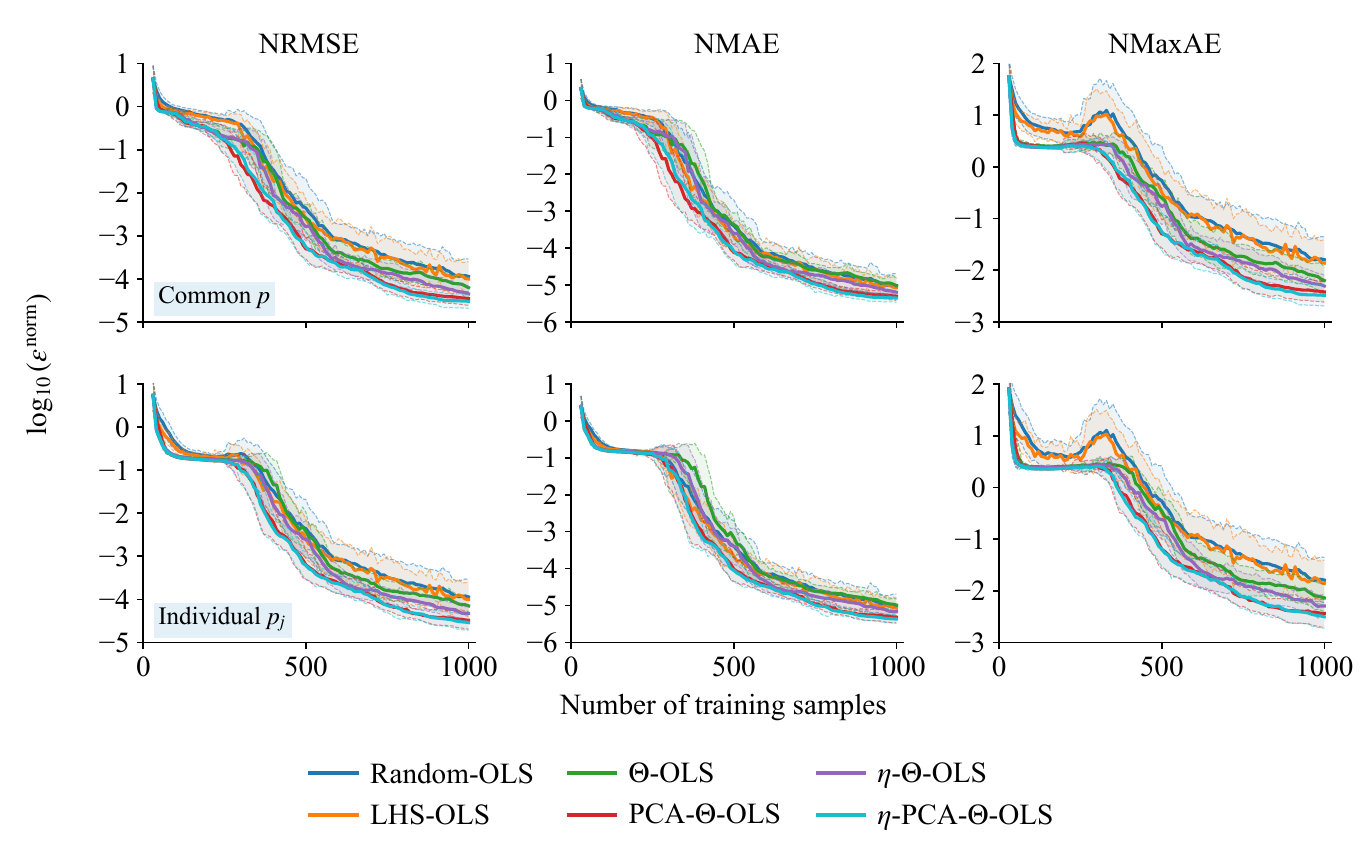}
    \caption{1D wave equation: error convergence (mean $\pm$ one standard deviation) with common degrees (top row) and individual degrees (bottom row).}
    \label{fig:wave_convergence}
\end{figure}

The final error rankings are similar under the two degree-selection rules (Fig.~\ref{fig:wave_convergence}). Within both pairs of adaptive methods, PCA reduces all three final test errors. The lowest final NRMSE, NMAE, and NMaxAE are obtained with $\eta$-PCA-$\Theta$-OLS under each rule. In particular, its final NMaxAE is approximately $3.3\times10^{-3}$ with a common degree and $3.1\times10^{-3}$ with individual degrees. For conventional $\Theta$-OLS, the corresponding values are $6.3\times10^{-3}$ and $7.3\times10^{-3}$. Therefore, individual degrees give a small improvement for the PCA-based criteria in this example, while conventional $\Theta$-OLS performs slightly better with a common degree.

\begin{figure}[htbp!]
    \centering
    \includegraphics[width=.93\linewidth]{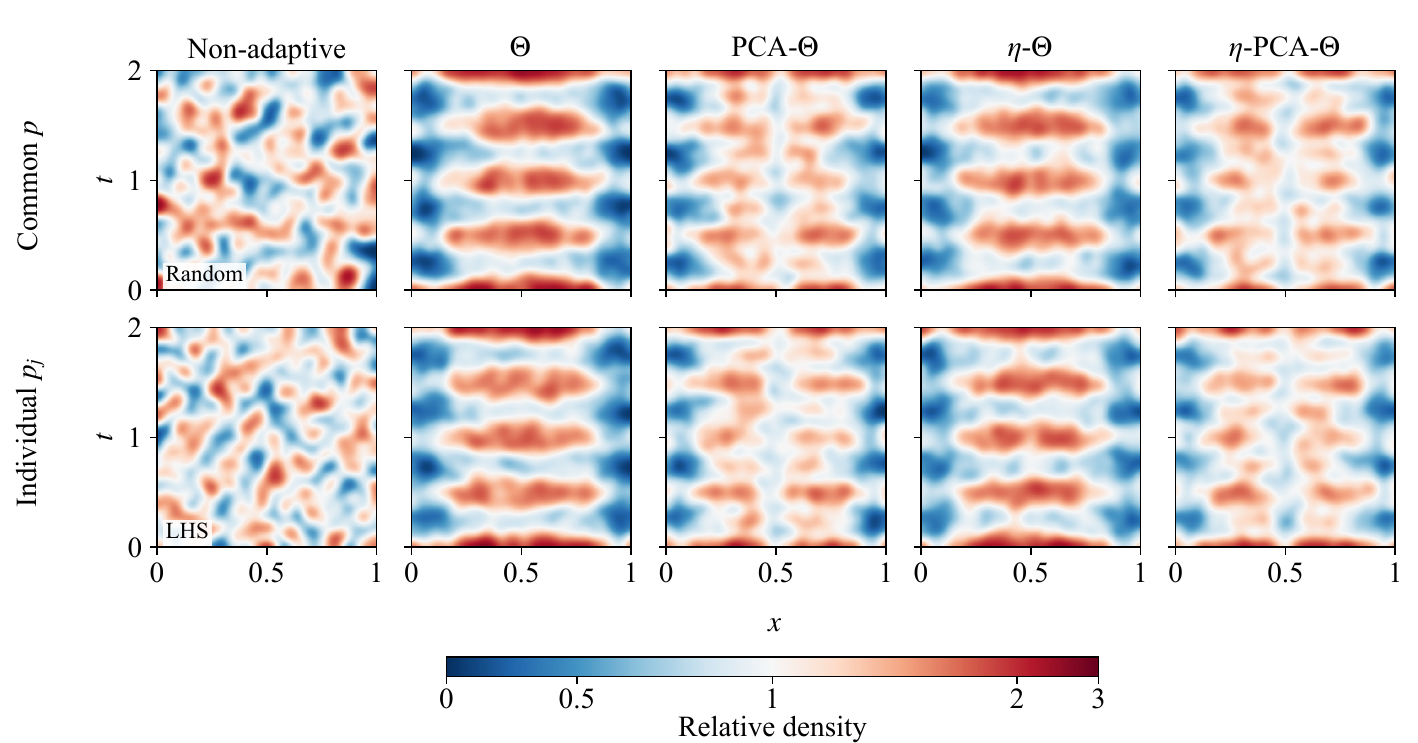}
    \caption{1D wave equation: sampling densities at $N=1000$ in a representative repetition. The first column shows the two non-adaptive designs, which are shared between degree-selection rules. The other columns compare common and individual degrees for each adaptive sampling method. All panels share a nonlinear colour scale with uniform density equal to one for visualization.}
    \label{fig:wave_sampling}
\end{figure}

The final sampling densities reflect the different spatial and temporal frequencies of the two modes (Fig.~\ref{fig:wave_sampling}). Conventional $\Theta$-OLS concentrates samples in bands near the extrema of the primary mode, particularly around $x=0.5$. With PCA, more samples are placed near the secondary-mode extrema around $x=0.25$ and $x=0.75$. These differences are visible under both degree-selection rules. Adaptive weighting moderates the concentration within the bands while preserving their modal structure due to the relatively uniform variance contribution of this example.

The previous and this examples illustrate how the exponent responds to the concentration of the variance contribution (Table~\ref{tab:wave_gaussian_weighting}). For a comparison, we use common-degree LOO selection in both examples and compare each weighted method with its own unweighted version. Because the Gaussian peaks concentrate the variance contribution in relatively limited regions, the mean exponent lies between $0.21$ and $0.28$. This compresses differences between values of two terms and gives the distance term greater influence on point selection. At $N=1000$, weighting reduces the fill distance $d_{fill}$, measured as the largest distance from an evaluation-grid point to its nearest training sample, by more than half for both criteria. In the wave example, the oscillatory features extend across the domain and the mean exponent is closer to one. Weighting then reduces the fill distance by only $7\%$ to $9\%$, with smaller gains in prediction accuracy. Additionally, at the final budget, both PCA-based methods retain four components in every repetition under either degree-selection rule. The corresponding saving in mean computational time is modest. For the conventional and adaptively weighted criteria, respectively, the reductions are $2.4\%$ and $3.0\%$ with a common degree, while $0.9\%$ and $1.9\%$ with individual degrees.

\begin{table}[!h]
    \centering
    \caption{Effect of adaptive weighting under common-degree LOO selection. The mean exponent is taken over all additions and repetitions. Reductions at $N=1000$ are relative to the corresponding unweighted criterion and are computed from average values over 50 repetitions.}
    \label{tab:wave_gaussian_weighting}
    \begin{tabular}{@{}llccccc@{}}
        \hline
        & & & \multicolumn{4}{c}{Reduction with weighting (\%)} \\
        \cline{4-7}
        Example & Criterion & Mean $\eta$ & $d_{fill}$& NRMSE & NMAE & NMaxAE \\
        \hline
        Gaussian & $\Theta$     & 0.208 & 57.6 & 95.3 & 96.1 & 82.9 \\
        Gaussian & PCA-$\Theta$ & 0.281 & 55.3 & 89.0 & 90.6 & 73.8 \\
        Wave     & $\Theta$     & 0.732 &  9.3 & 28.1 & 33.3 & 21.8 \\
        Wave     & PCA-$\Theta$ & 0.832 &  6.7 & 14.1 & 11.5 & 14.8 \\
        \hline
    \end{tabular}
\end{table}

\subsection{Multivariate random fields}
\label{sec:rf_example}

The final example uses multivariate RFs with 20 outputs to assess the proposed method in the presence of heterogeneous cross-correlations among the responses. Two complementary settings are examined. The first uses a two-dimensional trigonometric expansion with prescribed coefficient correlations to assess how consistently the sampling strategies perform across different field realizations. The second employs a Karhunen--Loève (KL) expansion over a five-dimensional input domain to compare the cost of evaluating the $\Theta$ criterion with and without PCA when both the polynomial basis and the candidate set are large.

\subsubsection{Two-dimensional random fields with prescribed cross-correlation and trigonometric bases}
\label{sec:rf_2d}

The first setting defines 20 response fields over the input domain $[0,1]^2$, which are divided into groups with different levels of within-group and between-group correlation. This construction produces strongly correlated outputs within each group while preserving distinct response information across groups. The prescribed trigonometric basis provides an explicit and controlled construction of the output fields with the desired modal complexity.

Each response field is constructed from a finite expansion of trigonometric modes,
\begin{equation}
F_j(\boldsymbol{x})
=
\sum_{k=1}^{K_{\mathrm{RF}}}
\sqrt{w_k}\,a_{kj}\phi_k(\boldsymbol{x}),
\qquad
j=1,\ldots,20,
\end{equation}
where $\phi_k$ denotes a spatial mode, $w_k$ is its normalized weight, and $a_{kj}$ is the coefficient associated with the $k$th mode and the $j$th output. The spatial modes are defined as
\begin{equation}
\begin{aligned}
\phi_{mn}^{\mathrm{ss}}(\boldsymbol{x})
&=
\sin(m\pi x_1)\sin(n\pi x_2),\\
\phi_{mn}^{\mathrm{cs}}(\boldsymbol{x})
&=
\cos(m\pi x_1)\sin(n\pi x_2),\\
\phi_{mn}^{\mathrm{sc}}(\boldsymbol{x})
&=
\sin(m\pi x_1)\cos(n\pi x_2),
\end{aligned}
\qquad
m,n=1,\ldots,5.
\end{equation}
The three mode series give a total of $K_{\mathrm{RF}}=3\times5\times5=75$
spatial modes. Their normalized weights are defined as
\begin{equation}
w_k
=
\frac{
\exp\!\left[-\gamma\left(m_k^2+n_k^2\right)\right]
}{
\displaystyle
\sum_{\ell=1}^{K_{\mathrm{RF}}}
\exp\!\left[-\gamma\left(m_\ell^2+n_\ell^2\right)\right]
},
\qquad
\gamma=0.1.
\end{equation}
The weights decrease with spatial frequency. The relatively slow decay for $\gamma=0.1$ retains contributions from a broad range of frequencies.

Dependence among the 20 outputs is introduced through the coefficient matrix of trigonometric modes
\begin{equation}
\mathbf{A}
=
\mathbf{Z}\mathbf{C}_{\mathrm{tar}}^{1/2},
\label{eq:rf_coefficient_transformation}
\end{equation}
where $\mathbf{Z}\in\mathbb{R}^{K_{\mathrm{RF}}\times20}$ contains independent standard normal entries and $\mathbf{C}_{\mathrm{tar}}^{1/2}$ is the symmetric square root of the prescribed correlation matrix. For each spatial mode $k$, the coefficient vector $\mathbf{a}_k=[a_{k1},\ldots,a_{k20}]$ has the prescribed correlation matrix $\mathbf{C}_{\mathrm{tar}}$.

\begin{figure}[!t]
    \centering
    \includegraphics[width=0.88\linewidth]{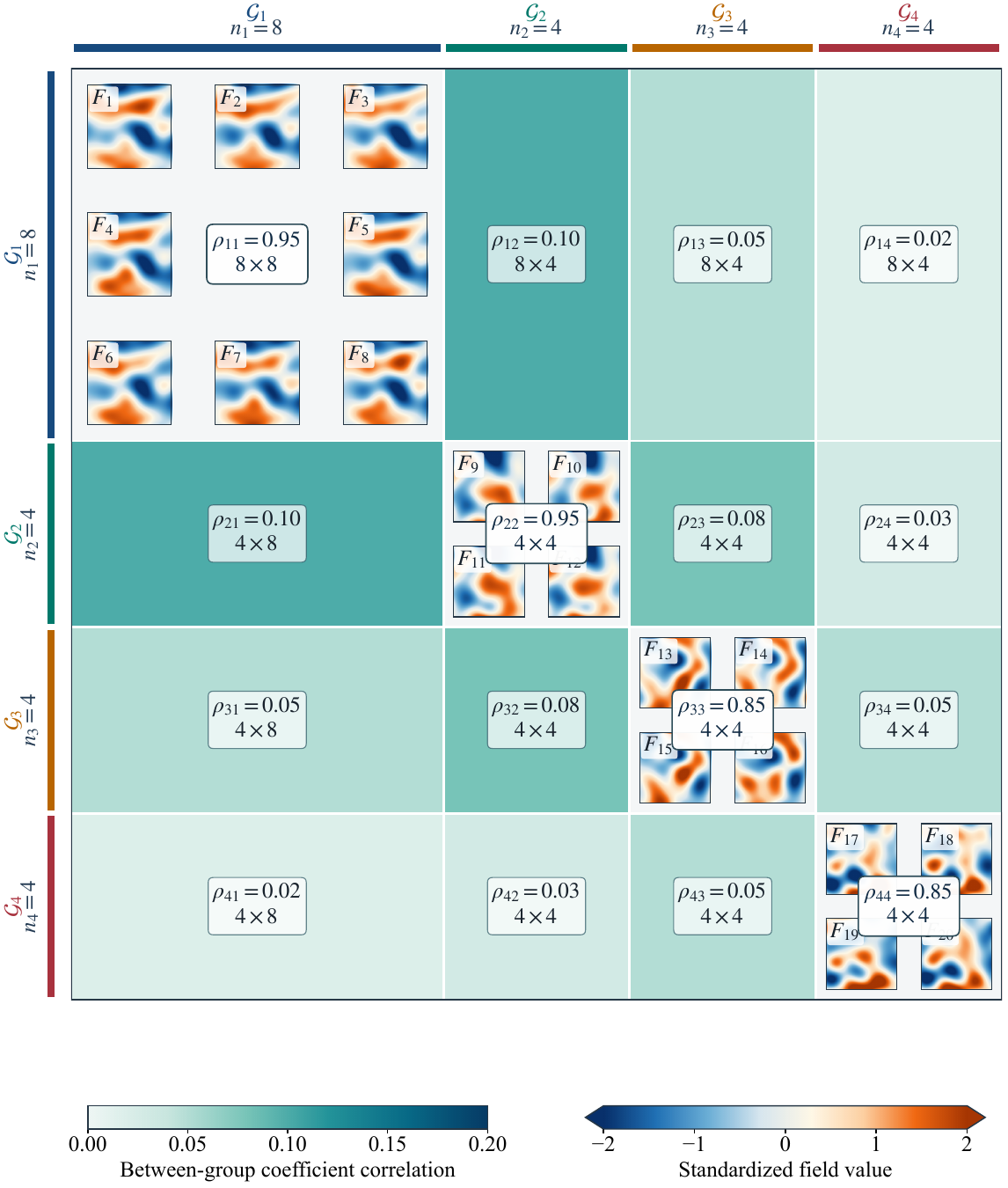}
    \caption{2D RFs with prescribed bases: prescribed blockwise coefficient-correlation structure and representative standardized output fields.}
    \label{fig:rf_matrix}
\end{figure}

The outputs are divided into four groups:
\begin{equation}
\mathcal{G}_1=\{1,\ldots,8\},
\qquad
\mathcal{G}_2=\{9,\ldots,12\},
\qquad
\mathcal{G}_3=\{13,\ldots,16\},
\qquad
\mathcal{G}_4=\{17,\ldots,20\}.
\end{equation}
The diagonal entries of $\mathbf{C}_{\mathrm{tar}}$ are equal to one. For two distinct outputs $i\in\mathcal{G}_g$ and $j\in\mathcal{G}_h$, the corresponding coefficient correlation is prescribed as
\begin{equation}
\left(\mathbf{C}_{\mathrm{tar}}\right)_{ij}
=
B_{gh},
\qquad i\neq j,
\end{equation}
where
\begin{equation}
\mathbf{B}
=
\begin{bmatrix}
0.95 & 0.10 & 0.05 & 0.02\\
0.10 & 0.95 & 0.08 & 0.03\\
0.05 & 0.08 & 0.85 & 0.05\\
0.02 & 0.03 & 0.05 & 0.85
\end{bmatrix}.
\end{equation}
The diagonal entries of $\mathbf{B}$ specify the correlations between distinct outputs within the same group, whereas the off-diagonal entries specify the correlations between outputs from different groups. Thus, $\mathcal{G}_1$ and $\mathcal{G}_2$ have the strongest within-group correlations, while $\mathcal{G}_3$ and $\mathcal{G}_4$ retain slightly more output variation. The relatively weak between-group correlations preserve distinct response patterns across the four groups. The prescribed correlations characterize dependence among the random modal coefficients over repeated realizations. They therefore need not coincide exactly with the empirical spatial correlations calculated between two output fields from a single realization. The prescribed blockwise coefficient-correlation structure and representative standardized output fields are shown in Fig.~\ref{fig:rf_matrix}.

The six sampling methods are compared over 50 repetitions using common-degree LOO selection. Each repetition uses a new field realization and a shared initial ED of 30 points. The training budget increases in steps of 10 to $N=1000$, with a $100\times100$ candidate grid for adaptive sampling and a common set of 1000 independent test points. The errors fluctuate as the degree increases and decrease more smoothly once it reaches its upper bound. The four adaptive methods generally achieve lower errors than Random-OLS and LHS-OLS during the early and intermediate stages of enrichment. The advantage is particularly clear in NMaxAE. At $N=1000$, all six sampling strategies achieve broadly comparable errors across the three measures.

\begin{figure}[!t]
    \centering
    \includegraphics[width=0.85\linewidth]{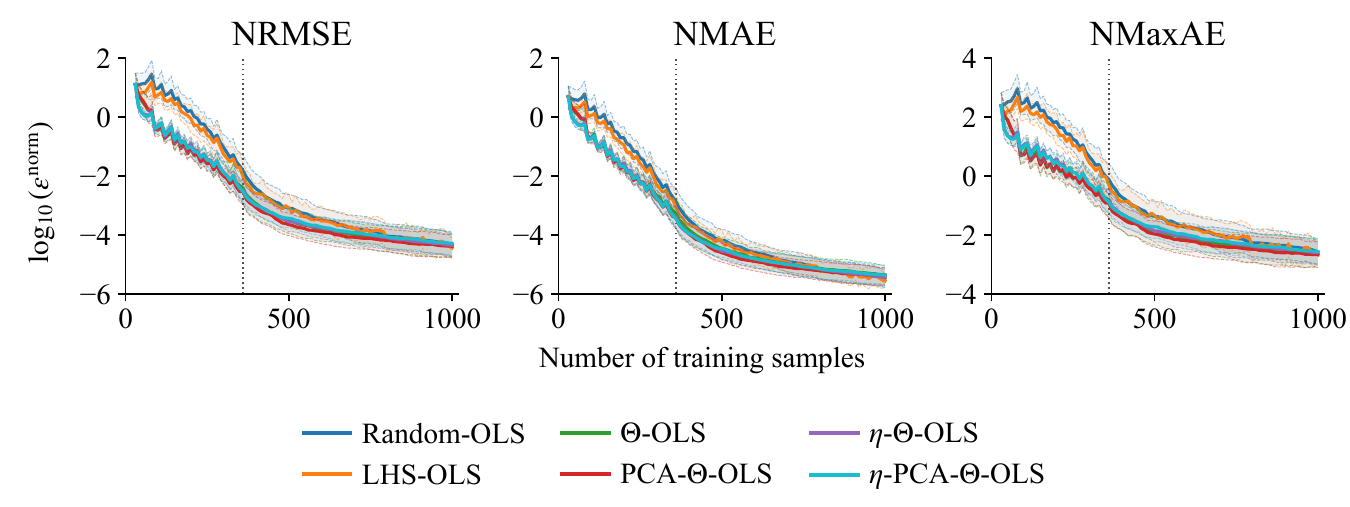}
    \caption{2D RFs with prescribed bases: convergence of the normalized prediction errors with common polynomial degree selected by LOO error (mean $\pm$ one standard deviation in logarithmic space).}
    \label{fig:rf_convergence}
\end{figure}

\begin{figure}[htbp!]
    \centering
    \includegraphics[width=0.85\linewidth]{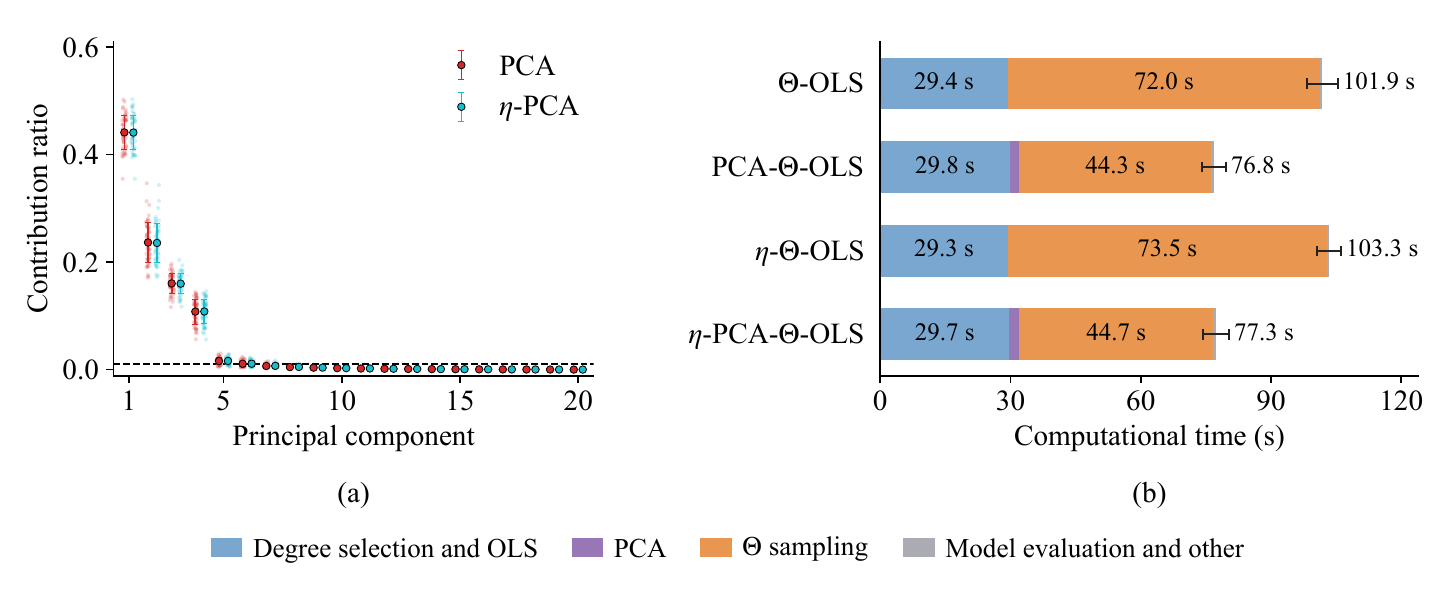}
    \caption{2D RFs with prescribed bases: (a) final PCA contribution ratios for the two PCA-based methods. The dashed line marks $r_{\min}=0.01$ and (b) mean computational time for the four adaptive methods.}
    \label{fig:rf_pca}
\end{figure}

At the final budget, the first four principal components account for about $95\%$ of the total standardized sample variance on the final ED for both PCA-based methods (Fig.~\ref{fig:rf_pca}a). With $r_{\min}=0.01$, five or six components are retained in most repetitions. Evaluating the criterion in this smaller response space reduces the mean computational cost from $101.9$ to $76.8$~s for the unweighted criterion and from $103.3$ to $77.3$~s with adaptive weighting, a saving of approximately $25\%$ in both cases (Fig.~\ref{fig:rf_pca}b). Specifically, most of this saving comes from evaluating local variance contributions for fewer response components. On average, PCA approximately adds $2$~s per run but saves around $25$~s with and without weighting overall.

\subsubsection{Five-dimensional random fields with prescribed cross-correlation and KL expansion}
\label{sec:rf_5d}

The second setting increases the input dimension to five and uses a KL expansion to generate 20 response fields over $[0,1]^5$. The underlying separable squared exponential covariance kernel is
\begin{equation}
C(\boldsymbol{x},\boldsymbol{x}')
=
\prod_{i=1}^{5}
\exp\!\left[
-\frac{(x_i-x_i')^2}{2\ell^2}
\right],
\qquad
\ell=0.4.
\end{equation}
The one-dimensional KL eigenproblems are solved using an 80-point Gauss-Legendre discretization, and the corresponding five-dimensional eigenpairs are formed by tensor products. Up to eight one-dimensional modes are considered in each coordinate, and the tensor-product modes are ordered by decreasing eigenvalue. Modes are retained until their cumulative contribution reaches $99\%$ of the total variance, resulting in $K_{\mathrm{KL}}=262$ retained modes. The response fields are constructed using these eigenfunctions and eigenvalues in place of the prescribed trigonometric modes and weights employed in the preceding setting. Cross-correlation is introduced through the same coefficient transformation given in Eq.~\eqref{eq:rf_coefficient_transformation}. The prescribed correlation matrix follows the same block structure as in the previous setting, except that all between-group correlations are set to \(0.05\). To limit the polynomial basis size and keep the required initial ED manageable in five dimensions, the polynomial degree is fixed at $p=10$ for all outputs and sampling methods. This results in $P=\binom{5+10}{10}=3003$ total-degree basis terms. The adaptive calculations use ordinary least squares regression and proceed from $N=3010$ to $N=8000$, with 10 samples added at each enrichment step. A common LHS candidate set of $2\times10^5$ points and an independent test set of 5000 points are used. For a consistent comparison, all methods use the same field realization, initial ED, and candidate set.

\begin{figure}[!h]
    \centering
    \includegraphics[width=0.95\linewidth]{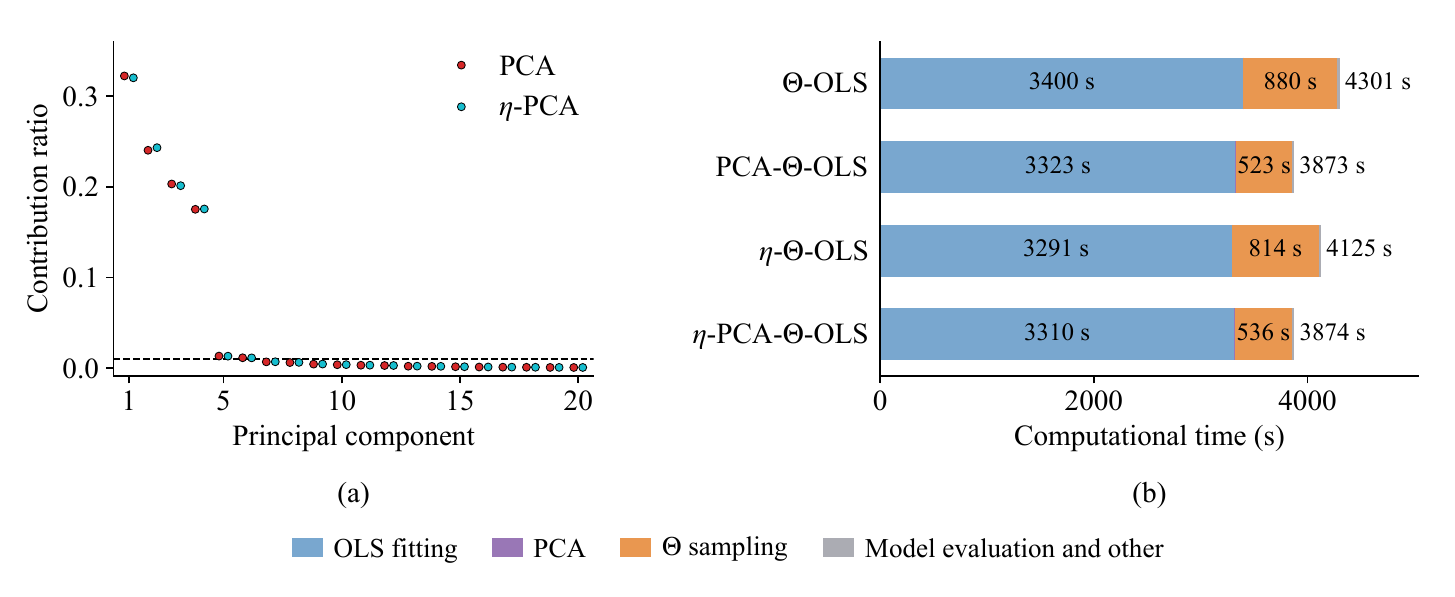}
\caption{5D RFs with KL expansion: (a) final PCA contribution ratios with threshold $r_{\min}=0.01$; (b) computational time for four adaptive methods.}
    \label{fig:kl5d_pca}
\end{figure}

Since the initial ED contains only seven more samples than the $3003$ basis terms, the initial OLS design matrix is checked by singular value decomposition. It has full numerical column rank and a two-norm condition number of $9.57\times10^4$. The conditioning improves rapidly for all four methods. After only 490 additional samples ($N=3500$), the condition numbers had fallen to below $1.05\times10^3$, before half of the total budget of 8000 samples used. At $N=6000$, the condition number has improved significantly and is generally less than $1\times10^2$. On the independent test set, the mean NRMSE across the 20 outputs decreases from 17.0\% for the shared initial fit to approximately $0.4\%$ at the final budget. Before the final enrichment step, both PCA-based methods retain six components at $r_{\min}=0.01$, with the first four accounting for approximately $94\%$ of the total standardized sample variance on the current ED (Fig.~\ref{fig:kl5d_pca}a).
Evaluating the $\Theta$ criterion over fewer response directions reduces the computational time spent on sampling by 40.5\% without adaptive weighting and by 34.1\% with it (Fig.~\ref{fig:kl5d_pca}b). Since PCA and adaptive weighting are used only to select new samples while all four methods fit PCEs with the 20 original outputs, OLS fitting takes a similar amount of time across methods and accounts for most of the computational cost because of the repeat of larger regression matrix. 

\section{Discussion \& Further Work}
\label{sec:discussion}

The proposed method aims to reduce the repeated influence of shared response features on sample selection. This is useful when the modeling objective is to resolve distinct features across the outputs, including those present in only a few QoIs. The choice of which principal components to retain also determines which response directions guide sampling. In the examples considered here, a few components generally accounted for most of the total standardized sample variance on the corresponding ED, and a fixed minimum contribution ratio of $r_{\min}=0.01$ provided a simple truncation rule. In contrast, when several components have comparable contributions, a cumulative variance threshold can be used to control the total retained fraction, or all numerically nonzero components can be retained. Additionally, close eigenvalues may make the corresponding component directions sensitive to small changes in the ED. Because each component contribution is normalized separately, a rotation of these directions can change the sampling order, although different sampling sequences may ultimately yield comparable prediction accuracy. In a separate sensitivity test at one enrichment stage of the wave example, small perturbations of the weights assigned to the ED points change the next selected point in two of 20 trials. Normalizing the contributions of the two components jointly gives a consistent selection across all 20 trials.

A further limitation arises from the linear nature of PCA. Outputs may exhibit weak linear correlation even when they are strongly related through a nonlinear mapping. Thus, future work could explore nonlinear representations, such as kernel PCA \cite{scholkopf1998nonlinear} or diffusion maps \cite{KONTOLATI2022111313}, to construct the sampling criterion. Within the present linear framework, dimensionality reduction could also be extended from sample selection to surrogate construction by fitting PCEs to the retained component scores and mapping their predictions back to the original output space. The full PCA transformation is invertible, whereas truncation can introduce a reconstruction error that must be considered alongside the PCE approximation error. This approach requires tracking the retained subspace during sequential enrichment and selecting its dimension based on both reconstruction and prediction accuracy. A further application is physics-constrained PCE (PC$^2$) \cite{novak2024physics,LU2026104001}. In this setting, the physical constraints and the associated Lagrange multiplier formulation would need to be expressed consistently in the reduced space so that the reconstructed outputs satisfy the governing equations and boundary conditions.

\section{Conclusion}
\label{sec:conclusion}

In this work, we propose a correlation-aware extension of the multi-output $\Theta$ criterion for adaptive PCE construction. At each enrichment stage, PCA identifies response directions from the empirical correlation structure of the standardized model responses on the current ED. Component-wise normalization and retention of selected directions then adjust their relative influence on sample selection, reducing the repeated weighting of shared response patterns. The distance-based exploration term and the final PCE surrogates for the original outputs remain unchanged. For problems with localized response features, adaptive weighting helps maintain coverage of the input domain by moderating the influence of concentrated variance contributions on sample selection. The numerical results show that the proposed method maintains the accuracy and convergence behavior of the conventional multi-output \(\Theta\) method while providing a more compact representation of correlated responses. By reducing the number of response directions involved in the sampling criterion, PCA can also reduce the computational cost, with greater benefits for problems involving many strongly correlated outputs and large polynomial bases or candidate sets. Overall, the proposed PCA-based criterion provides an efficient and practical approach to correlation-aware active learning for multi-output PCE surrogate modeling of complex physical systems.

\section*{Data Availability}

The data and scripts that support the findings of this study are openly available at \url{https://github.com/QitianLu/PCA_Theta}.

\section*{Acknowledgments}

The work of L.N. was supported by the Czech Science
Foundation under project No.~26-20456S.
The work of M.V. was supported by the European Union's Horizon
Europe research and innovation programme through the European
Research Council (ERC) Synergy Grant FatResCon
(grant agreement No.~101167045).

\begin{sloppypar}
   
\printbibliography
\end{sloppypar}

\end{document}